\documentclass[fleqn,usenatbib]{mnras}

\usepackage{newtxtext,newtxmath}

\usepackage[T1]{fontenc}
\usepackage{multirow}

\DeclareRobustCommand{\VAN}[3]{#2}
\let\VANthebibliography\thebibliography
\def\thebibliography{\DeclareRobustCommand{\VAN}[3]{##3}\VANthebibliography}

\usepackage{graphicx}	
\usepackage{amsmath}	
\usepackage{multirow}
\usepackage[T1]{fontenc}

\title[Predicting reliable H$_2$ column density]{Predicting reliable H$_2$ column density maps from molecular line data using machine learning}

\author[Y. Shimajiri et al.]{
Yoshito SHIMAJIRI,$^{1}$\thanks{E-mail: y-shimajiri@fains.jp}
Yasutomo KAWANISHI,$^{2}$
Shinji FUJITA,$^{3,4}$
Yusuke MIYAMOTO,$^{5}$
Atsushi M. ITO,$^{6}$\newauthor
Doris ARZOUMANIAN,$^{7}$
Philippe ANDR$\acute{\rm E}$, $^{8}$
Atsushi NISHIMURA,$^{7}$
Kazuki TOKUDA,$^{4,7,9}$
Hiroyuki KANEKO,$^{7,10}$\newauthor
Shunya TAKEKAWA,$^{11}$
Shota UEDA,$^{4}$
Toshikazu ONISHI,$^{4}$
Tsuyoshi INOUE$^{12}$
Shimpei NISHIMOTO,$^{4}$\newauthor
and
Ryuki YONEDA$^{4}$
\\
$^{1}$ Kyushu Kyoritsu University, Jiyugaoka 1-8, Yahatanishi-ku, Kitakyushu, Fukuoka, 807-08585, Japan\\
$^{2}$ RIKEN Information R\&D and Strategy Headquarters, 2-2-2 Hikaridai, Seika-cho, Soraku-gun, Kyoto, 619-0288, Japan\\
$^{3}$ Institute of Astronomy, Graduate School of Science, The University of Tokyo, 2-21-1 Osawa, Mitaka, Tokyo 181-0015, Japan \\
$^{4}$ Department of Physics, Graduate School of Science, Osaka Metropolitan University, 1-1 Gakuen-cho, Naka-ku, Sakai, Osaka 599-8531, Japan\\
$^{5}$Department of Electrical and Electronic Engineering, Fukui University of Technology, 3-6-1, Gakuen, Fukui, 910-8505, Japan\\
$^{6}$ National Institute for Fusion Science (NIFS), National Institutes of Natural Sciences (NINS), 322-6, Oroshi-cho, Toki, Gifu, 509-5292, Japan\\
$^{7}$ National Astronomical Observatory of Japan, Osawa 2-21-1, Mitaka, Tokyo, 181-8588, Japan\\
$^{8}$ Laboratoire d'Astrophysique (AIM), Universit\'e Paris-Saclay, Universit\'e Paris Cit\'e, CEA, CNRS, AIM, 91191 Gif-sur-Yvette, France \\
$^{9}$ Department of Earth and Planetary Sciences, Faculty of Science, Kyushu University, Nishi-ku, Fukuoka 819-0395, Japan\\
$^{10}$ Graduate School of Education, Joetsu University of Education, 1, Yamayashiki-machi,Joetsu, Niigata 943-8512, Japan\\
$^{11}$ Department of Applied Physics, Faculty of Engineering, Kanagawa University, 3-27-1 Rokkakubashi, Kanagawa-ku,Yokohama, Kanagawa, 221-8686, Japan\\
$^{12}$ Department of Physics, Konan University, 8-9-1 Okamoto, Higashinada-ku, Kobe, Hyogo, 658-8501, Japan
}

\date{Accepted XXX. Received YYY; in original form ZZZ}

\pubyear{2015}

\begin{document}
\label{firstpage}
\pagerange{\pageref{firstpage}--\pageref{lastpage}}
\maketitle

\begin{abstract}
The total mass estimate of molecular clouds suffers from the uncertainty in the H$_2$-CO conversion factor, the so-called $X_{\rm CO}$ factor, which is used to convert the $^{12}$CO (1--0) integrated intensity to the H$_2$ column density. We demonstrate the machine learning’s ability to predict the H$_2$ column density from the $^{12}$CO, $^{13}$CO, and C$^{18}$O (1--0) data set of four star-forming molecular clouds; Orion A, Orion B, Aquila, and M17. When the training is performed on a subset of each cloud, the overall distribution of the predicted column density is consistent with that of the {\it Herschel} column density. The total column density predicted and observed is consistent within 10\%, suggesting that the machine learning prediction provides a reasonable total mass estimate of each cloud. However, the distribution of the column density for values $> \sim 2 \times 10^{22}$ cm$^{-2}$, which corresponds to the dense gas, could not be predicted well. This indicates that molecular line observations tracing the dense gas are required for the training. We also found a significant difference between the predicted and observed column density when we created the model after training the data on different clouds. This highlights the presence of different $X_{\rm CO}$ factors between the clouds, and further training in various clouds is required to correct for these variations. We also demonstrated that this method could predict the column density toward the area not observed by {\it Herschel} if the molecular line and column density maps are available for the small portion, and the molecular line data are available for the larger areas.
\end{abstract}



\begin{table*}
\footnotesize
\centering  
\caption{Data-set \label{table:data}}  
\begin{tabular}{lcccc}   
\hline   
Cloud            &  Orion A   & Orion~B/NGC~2024   & Aquila & M~17 \\
\hline
$^{12}$CO(1-0)$^{\dag}$ &   \citet{Shimajiri11,Nakamura19}   &  \citet{Shimajiri2023}    &  \citet{Shimoikura19aquila}  & \citet{Shimoikura19,Sugitani19} \\
$^{13}$CO(1-0)$^{\dag}$ &  \citet{Shimajiri11,Shimajiri14}   &  \citet{Shimajiri2023}    &  \citet{Shimoikura19aquila}  & \citet{Shimoikura19,Sugitani19} \\
C$^{18}$O(1-0)$^{\dag}$ &  \citet{Shimajiri11,Shimajiri14}  &  \citet{Shimajiri2023}    &  \citet{Shimoikura19aquila} & \citet{Shimoikura19,Sugitani19}\\
{$N_{\rm H_2}$}$^{\dag}$ &   \citet[]{Andre10}$^{\ddag}$, Pezzuto et al. in prep.  &  \citet{Konyves20}$^{\ddag}$    & \citet{Konyves15}$^{\ddag}$  &  \citet{Motte10}$^{\spadesuit}$ \\
\hline   
\multicolumn{5}{l}{[$^{\dag}$] $^{12}$CO(1--0), $^{13}$CO(1--0), and C$^{18}$O(1--0) data are obtained with the Nobeyama 45m. The column density $N_{\rm H_2}$ is derived from {\it Herschel} data.}\\
\multicolumn{5}{l}{[$^{\ddag}$] HGBS Archive (\url{http://gouldbelt-herschel.cea.fr/archives}).}\\
\multicolumn{5}{l}{[$^{\spadesuit}$] HOBYS Archive (\url{http://hobys-herschel.cea.fr}).}
\end{tabular}   
\end{table*}

\section{Introduction}\label{sect:introduction}

In astronomy, especially in studies of the interstellar medium (ISM), it is essential to investigate the spatial distribution and mass of molecular hydrogen, H$_2$, since molecular hydrogen is the most abundant molecule in the universe. Since it is difficult to directly observe the H$_2$ emission in the cold ($\sim$10--20K) ISM, the $^{12}$CO (1--0) emission is instead used to measure the mass of the molecular gas by using the relation between the $^{12}$CO integrated intensity and the H$_2$ column density of $N({\rm H_2}) {\rm [cm^{-2}]} = X_{\rm CO} {\rm [cm^{-2}/K\ km\ s^{-1}]} \times W(^{12}{\rm CO}) {\rm [K\ km\ s^{-1}]}$ where $N({\rm H_2})$, $W(^{12}{\rm CO})$, and $X_{\rm CO}$ are the H$_2$ column density, the integrated intensity of $^{12}$CO (1--0), and H$_2$-to-CO conversion factor, respectively. \citet{Bolatto13} found a conversion factor of $X_{\rm CO}$ = 2$\times$10$^{20}$ ${\rm [cm^{-2}/K\ km\ s^{-1}]}$ with $\pm$30\% uncertainty in the Milky Way. The column density is also derived from the molecular line data assuming the local thermodynamic equilibrium (LTE) condition by the following equation \citep[e.g.,][]{Mangum15}.

\begin{equation}\label{Nmol}
\centering
N_{\rm mol} = \frac{3h}{8\pi^3} \frac{Q}{\mu S_{i,j}} \frac{e^{E_{\rm u}/kT_{\rm ex}}}{e^{h\nu/kT_{\rm ex}}-1} \int \tau dv
\end{equation}

\noindent where $k, h, v, Q, \mu, E_{\rm u}, S_{\rm i,j}, {\rm and}\ \tau$ are the Boltzmann constant, the Planck constant, the rest frequency of the molecular line, the partition function, the dipole moment, the energy of the upper level, and the intrinsic line strength of the transition for $i$ to $j$ state, and the optical depth. $T_{\rm ex}$ is the excitation temperature which can be derived from $^{12}$CO (1--0) as follows \citep[c.f.,][]{Nishimura15}:

\begin{equation}\label{EQ:Tex}
\centering
T_{\rm ex} = 5.53 \{\ln [1+\frac{5.53}{T_{\rm peak}+0.84}]\}^{-1}
\end{equation}

\noindent where $T_{\rm peak}$ is the peak intensity of $^{12}$CO (1--0).
To convert the $N_{\rm mol}$ to H$_2$ column density $N({\rm H}_2)$, the molecular abundance $X_{\rm mol}$ is also required (i.e., $N({\rm H}_2) = N_{\rm mol}/X_{\rm mol}$). However, molecular abundance is known to have a large variation on a wide scale range from the core scale to the scale of a galaxy \citep[c.f.][]{Watanabe14, Shimajiri15b, Nishimura16, Shimajiri17, Watanabe19,Tokuda21}. Furthermore, the abundance ratios between $^{13}$CO and C$^{18}$O and between $^{12}$CO and $^{13}$CO change due to the selective FUV (far ultraviolet) dissociation, suggesting that these abundances vary with the FUV radiation \citep{Shimajiri14, Nishimura15, Lin16, Ishii19}. The FUV radiation changes the molecular abundance in the ISM. Thus, adopting the appropriate molecular abundance and taking into account the FUV radiation effect are crucial to derive the H$_2$ column density accurately.

As mentioned above, the mass estimate has a large uncertainty despite its importance in describing the physical properties of molecular clouds. The amount of information increases rapidly with the advent of astronomical instruments. 
Recently, machine learning methods have been applied to various astronomical data sets for diverse scientific objectives such as the classification of galaxies \citep{Barchi20}, clustering into exoplanets \citep{Schanche19}, identification of the ring structures surrounding the HII region \citep{Ueda20, Nishimoto22}, solving Near and far problem in inner Galaxy \citep{Fujita23}, identifying the filamentary structure \citep{Zavagno23} and so on.
\citet{Gratier21} demonstrated that the H$_2$ column density could be predicted from multi-molecular line emission using one of the machine-learning methods, the {\it random forest}. They used multi-molecular line data toward Orion~B obtained with the IRAM30m telescope. They found that the $^{12}$CO, $^{13}$CO, and C$^{18}$O lines play significant roles in predicting the H$_2$ column density. For higher-density areas, the dense gas tracers such as HNC, HCO$^{+}$, and N$_2$H$^+$ are also important.

This study uses the machine learning technique to predict the H$_2$ column density from the combined analysis of the $^{12}$CO, $^{13}$CO, and C$^{18}$O data sets. 
The $^{12}$CO, $^{13}$CO, and C$^{18}$O emissions are frequently used to derive the mass of the ISM and to trace the cloud structures well. With the advent of new receivers and correlators (such as those on the Nobeyama 45m telescope), it is  feasible to observe these lines simultaneously. Furthermore, the intensities of these lines are stronger than those of the dense gas tracers such as H$^{13}$CO$^{+}$, and N$_2$H$^+$. Consequently, wide-field mapping of molecular clouds is possible with these three CO isotopologues. In fact, many surveys toward star forming regions in $^{12}$CO, $^{13}$CO, and C$^{18}$O have been conducted \citep{Umemoto17,Braiding18,Nakamura19,Su19,Torii19}. Thus, revealing how accurately the structure and mass of molecular clouds can be reconstructed from the $^{12}$CO, $^{13}$CO, and C$^{18}$O using machine learning will provide new tools to study the ISM physics and star formation. 
This paper is organized as follows: In Sect. \ref{sec:data}, we describe the data set used in the paper. In Sect. \ref{sec:machine_learning}, we described the data preprocessing, optimization setup, training, and test. In Sect. \ref{sec:results}, we show the results derived from applying the model produced by the Extra Trees Regressor \citep[ET,][]{Geurts2006}. 
In Sect. \ref{sect:discuss}, we will evaluate the accuracy of column density predictions and predict column densities at higher spatial resolution. Our conclusions are summarized in Sect. \ref{sect:concle}.

\begin{figure*}
\centering
\includegraphics[angle=0,width=18cm]{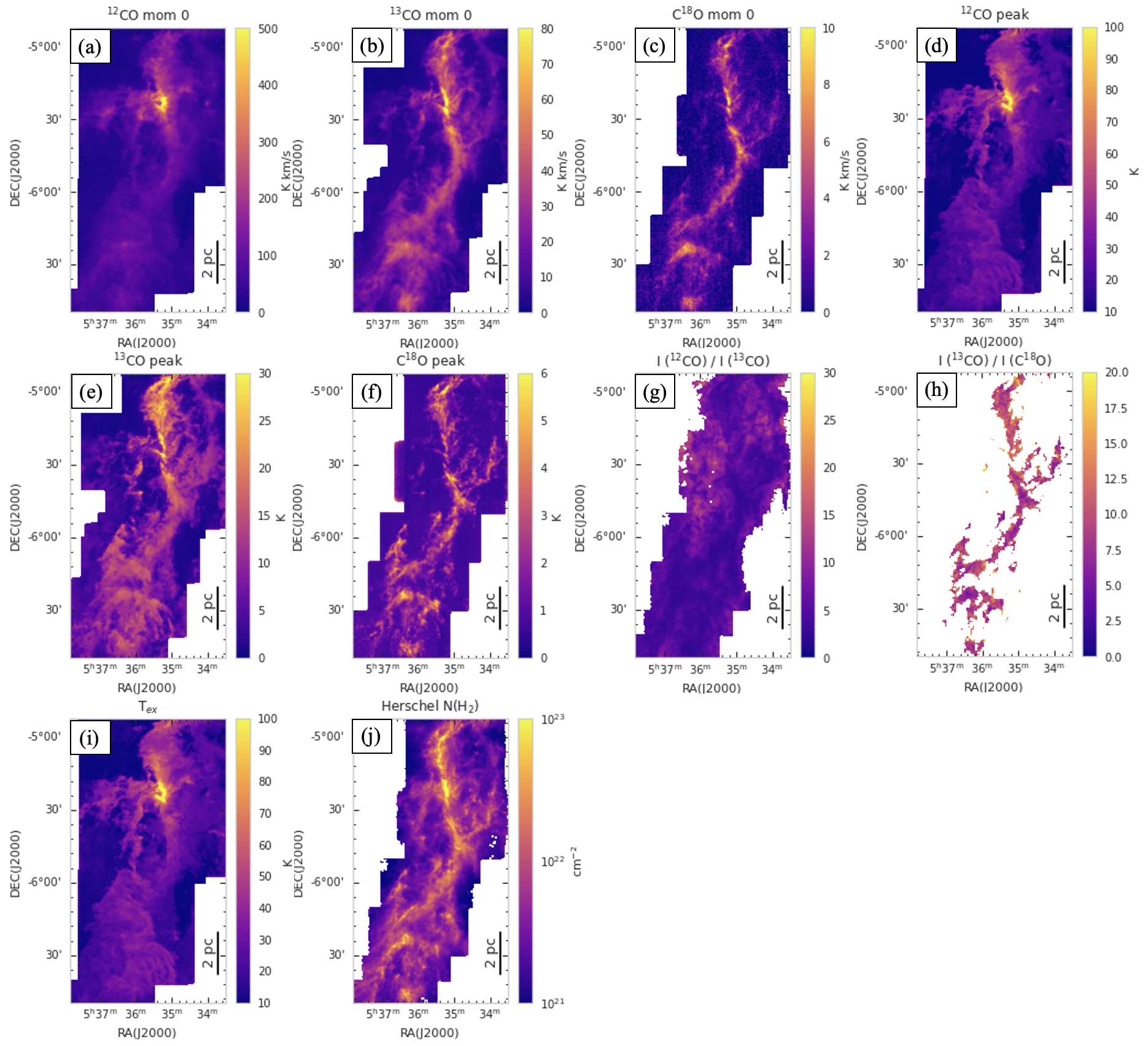}
\caption{($a-c$) $^{12}$CO (1--0), $^{13}$CO (1--0), and C$^{18}$O (1--0) integrated intensity (moment 0) maps, ($d-f$) $^{12}$CO, $^{13}$CO, and C$^{18}$O peak intensity maps, ($g-h$) maps of the integrated intensity ratio of $^{12}$CO to $^{13}$CO and of $^{13}$CO to C$^{18}$O, ($i$) excitation temperature map, and ($j$) HGBS H$_2$ column density map toward Orion~A molecular cloud. The angular resolution of all maps is smooth to 25$\arcsec$. 
}
\label{fig_orionA_obs}
\end{figure*}

\section{Data set}\label{sec:data}

Maps of $^{12}$CO ($J$=1--0), $^{13}$CO ($J$=1--0), and C$^{18}$O ($J$=1--0) toward Orion~A, Orion~B/NGC~2024, Aquila, and M~17 are publicly available from \cite{Shimajiri11,Shimajiri14,Shimoikura19,Shimoikura19aquila,Sugitani19,Nakamura19} (see Table \ref{table:data}). 

In parallel, good H$_2$ column density maps of the same regions derived from {\it Herschel} submillimeter dust continuum data are available \citep[see Table \ref{table:data},][Pezzuto et al. in prep.]{Andre10,Motte10,Konyves15,Konyves20}.

\begin{table*}  
\caption{Example of the list used as input for the training}
\label{table:dataframe}
\scriptsize	
\centering
\begin{tabular}{lccccccccccccc}   
\hline   
 &	\multirow{2}{*}{RA$_{\rm pix}$} &	\multirow{2}{*}{DEC$_{\rm pix}$}	 & $^{12}$CO &	$^{13}$CO   &	C$^{18}$O  &	$^{12}$CO 	& $^{13}$CO   &	C$^{18}$O &	\multirow{2}{*}{$T_{\rm ex}$}	 & \multirow{2}{*}{$\frac{\rm ^{12}CO~mom0}{\rm ^{13}CO~mom0}$}  & \multirow{2}{*}{$\frac{\rm ^{13}CO~mom0}{\rm C^{18}O~mom0}$} &	\multirow{2}{*}{$N$(H$_2$)} & \multirow{2}{*}{cloud} \\
  &	 &	 &   Mom 0 &	 Mom 0 &	 Mom 0 &	 Peak	&  Peak &	  Peak & &   &  &	&  \\
  &	 &	 &   [K km/s] &	  [K km/s] & [K km/s] &	 [K]	&  [K]  &[K] & [K] &   &  & [$¥\times$10$^{21}$cm$^{-2}$]	&  \\  
 \hline
383	&145&	245&	46.027042&	13.917648&	1.410226&	17.584301&	7.415209&	1.891411&	21.047360&	3.307099&	9.869093&	5.025195&	orion\\
432&	146	&205&	53.787624&	14.213110&	0.627789&	13.870554&	8.023403&	2.044455&	17.307561&	3.784367&	22.639942&	8.674680&	orion\\
433&	146&	206&	55.570644&	15.812763&	0.706174&	14.270488&	7.973323&	2.304390&	17.710831&	3.514291&	22.392151&	10.559992&	orion\\
437&	146&	210&	62.206879&	20.084249&	0.945845&	15.013415&	9.388056&	1.692411&	18.459568&	3.097297&	21.234194&	12.494176&	orion\\
438&	146&	211&	63.578884&	20.229988&	0.972942&	15.296682&	9.360451&	2.014358&	18.744927&	3.142804&	20.792583&	12.344450&	orion\\
... &... 	& ... & ... &	...  & 	...  &...  &...  & ...  &...  &...  &...  &...  & 	... \\
\hline   
\multicolumn{14}{l}
{Example of the list used as input for the training 
with from left to right, a pixel running number, the coordinates in pixel, the integrated intensity, peak intensity, of $^{12}$CO, $^{13}$CO, and C$^{18}$O,}\\
\multicolumn{14}{l}
{the ratio of $^{12}$CO to $^{13}$CO, the ratio of $^{13}$CO to C$^{18}$O, the excitation temperature, the HGBS H$_2$ column density, and the cloud name.}
\end{tabular} 
\end{table*}

\begin{table*}  
\caption{\textcolor{black}{Tuned hyperparameters for each data-set}}
\label{table:model_tuned_param}
\centering
\begin{tabular}{lcccccc}   
\hline
Regressor  &  Regressor- & Regressor- & Regressor- & Regressor- & Regressor-  & Regressor-\\
 & OMC1 & OMC123  &  NGC2024 & Aquila & M17 & High-Resolution\\
\hline
area used for training & Orion~A/ & Orion~A/ &  Orion~B/ & Aquila & M~17 &  Orion~A \\
& OMC-1 & OMC-1/2/3& NGC~2024 & & &  OMC-1/2/3  \\
\hline
blc\_RA$^{\dag}$ & 5$^{\rm h}$36$^{\rm m}$8$^{\rm s}$.8 & 5$^{\rm h}$36$^{\rm m}$8$^{\rm s}$.8 & 5$^{\rm h}$42$^{\rm m}$14$^{\rm s}$.7 & 18$^{\rm h}$31$^{\rm m}$43$^{\rm s}$.1 & 0$^{\rm h}$57$^{\rm m}$22$^{\rm s}$.0 &  5$^{\rm h}$5$^{\rm m}$47$^{\rm s}$.4  \\
blc\_DEC$^{\dag}$ & -5$^{\rm d}$29$^{\rm m}$43$^{\rm s}$.7 & 5$^{\rm d}$33$^{\rm m}$28$^{\rm s}$.7 & -1$^{\rm d}$58$^{\rm m}$5$^{\rm s}$.9 & -2$^{\rm d}$12$^{\rm m}$29$^{\rm s}$.0 & 0$^{\rm d}$36$^{\rm m}$45$^{\rm s}$.0 & -5$^{\rm d}$29$^{\rm m}$44$^{\rm s}$.9 \\
trc\_RA$^{\ddag}$ & 5$^{\rm h}$34$^{\rm m}$28$^{\rm s}$.4 & 5$^{\rm h}$34$^{\rm m}$28$^{\rm s}$.4 & 5$^{\rm h}$41$^{\rm m}$27$^{\rm s}$.2 & 18$^{\rm h}$29$^{\rm m}$50$^{\rm s}$.0 & 0$^{\rm h}$55$^{\rm m}$22$^{\rm s}$.0 &  5$^{\rm h}$34$^{\rm m}$42$^{\rm s}$.3  \\
trc\_DEC$^{\ddag}$ & -5$^{\rm d}$15$^{\rm m}$58$^{\rm s}$.8 & -4$^{\rm d}$55$^{\rm m}$58$^{\rm s}$.8 &-1$^{\rm d}$54$^{\rm m}$58$^{\rm s}$.4 &-1$^{\rm d}$58$^{\rm m}$44$^{\rm s}$.2 & 0$^{\rm d}$26$^{\rm m}$44$^{\rm s}$.9 & -4$^{\rm d}$57$^{\rm m}$24$^{\rm s}$.8 \\
\hline
number of samples (pixels) & \multirow{2}{*}{22311} & \multirow{2}{*}{60501} & \multirow{2}{*}{2496} & \multirow{2}{*}{25197} & \multirow{2}{*}{19521} & \multirow{2}{*}{472269}\\
\hspace{5mm}in training set  & & & & & & \\
number of samples (pixels) & \multirow{2}{*}{330030} & \multirow{2}{*}{330030} & \multirow{2}{*}{25124} & \multirow{2}{*}{131040} & \multirow{2}{*}{106328} & \multirow{2}{*}{5426235} \\
\hspace{5mm}in the area observed in CO/$^{13}$CO/C$^{18}$O & & & & & & \\
\hline
max\_depth$^1$ & 11& 11 & 7 & 11 & 11 & 11\\
max\_features$^2$ & 1 & 1 & 1 & 1 &1 & 1\\
min\_samples\_leaf$^3$ & 1 & 1 & 2  & 1 &1 & 1 \\
min\_samples\_split$^4$ & 2& 2 & 5 & 3  & 3 & 5\\
min\_weight\_fraction\_leaf$^5$ & 0 & 0 & 0 & 0 & 0 & 0\\
n\_estimators$^{6}$ & 245 & 300 & 244 & 241 & 256 & 251 \\

\hline   
\multicolumn{6}{l}{$^{\dag}$: RA and DEC coordinates of the bottom left corner of the trained area.} \\
\multicolumn{6}{l}{$^{\ddag}$: RA and DEC coordinates of the top right corner of the trained area.} \\
\multicolumn{6}{l}{$^1$:The maximum depth of the tree.} \\
\multicolumn{6}{l}{$^2$:The limit of the maximum number of features used for each tree. }\\
\multicolumn{6}{l}{$^3$:The minimum number of samples required to be at a leaf node. }\\
\multicolumn{6}{l}{$^4$:The minimum number of samples required to split an internal node.}\\
\multicolumn{6}{l}{$^5$:The minimum weighted fraction of the sum total of weights (of all the input samples) required to be at a leaf node.}\\
\multicolumn{6}{l}{$^{6}$: The number of trees in the forest.}\\
\end{tabular}  
\end{table*}


The $^{12}$CO, $^{13}$CO, and C$^{18}$O data were obtained with the Nobeyama 45m telescope. The angular resolution of the data used in this study is 25$\arcsec$ smoothed from the original effective angular resolution of 20-21$\arcsec$ to improve the sensitivity. The integrated intensity and peak intensity maps are then produced from these cube data. 
The integrated intensity ratios of $^{12}$CO to $^{13}$CO and of $^{13}$CO to C$^{18}$O are also produced (see Fig. \ref{fig_orionA_obs}).

\begin{figure*}
\centering
\includegraphics[angle=0,width=14cm]{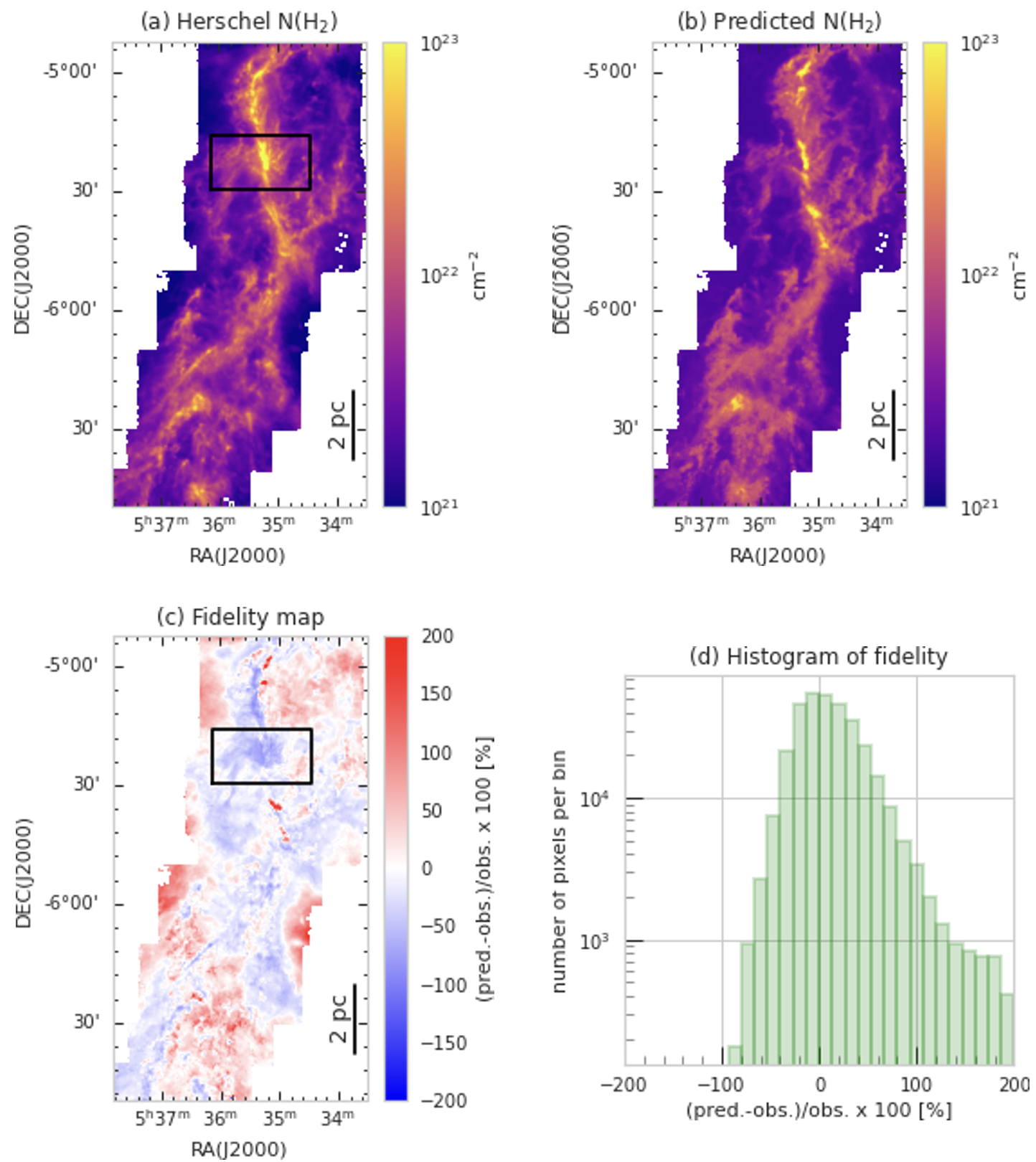}
\caption{Maps of ($a$) the HGBS H$_2$ column density, ($b$) the H$_2$ column density predicted by Regressor-OMC1,  ($c$) the fidelity between the observations and the prediction toward Orion~A molecular cloud, and ($d$) histogram of the fidelity. 
The black box in panels (a) and (c) indicates the area used for training the regressor. In panel ($b$), the H$_2$ column density is predicted by Regressor-OMC1 from the molecular line data.
}
\label{fig_OMC1_prediction}
\end{figure*}

\begin{table*}  
\footnotesize
\caption{\textcolor{black}{Comparison among models}}
\label{table:comp_model}
\centering  
\begin{tabular}{llrrrrrrr}
\hline
{} &                            Model &      MAE &       MSE &     RMSE &      R2 &   RMSLE &    MAPE &  TT  \\
{} &                             &      {\it Mean Absolute}  &      {\it Mean Square} &     {\it Root Mean}  &      {\it R-Squared} &   {\it Root Mean}  &    {\it Mean Absolute}  &  {\it Training}   \\
{} &                             &       {\it Error} &       {\it Error} &    {\it Square Error} &     &   {\it Square Error} &   {\it Percentage Error} &   {\it Time} \\
\hline
  & & [$\times$10$^{21}$ cm$^{-2}$]  &   &   [$\times$10$^{21}$ cm$^{-2}$]  &  &  & [\%] & [sec]\\
\hline
et       &            Extra Trees Regressor &   {\bf 2.3205} &    {\bf 92.2658} &   {\bf 8.8934} &  {\bf 0.9020} &  {\bf 0.1680} &  {\bf 0.1451} &     0.962 \\
lightgbm &  Light Gradient Boosting Machine &   3.2010 &   213.8056 &  13.0801 &  0.8457 &  0.2102 &  0.1899 &     0.599 \\
rf       &          Random Forest Regressor &   2.7379 &   151.9937 &  11.2724 &  0.8166 &  0.1827 &  0.1571 &     0.872 \\
gbr      &      Gradient Boosting Regressor &   3.3099 &   160.4642 &  12.0486 &  0.8001 &  0.2365 &  0.2304 &     0.567 \\
ada      &               AdaBoost Regressor &  10.0274 &   287.9909 &  16.4738 &  0.6675 &  0.8676 &  1.7285 &     0.188 \\
knn      &            K Neighbors Regressor &   4.8112 &   420.5454 &  19.3340 &  0.6282 &  0.2777 &  0.2324 &     0.135 \\
dt       &          Decision Tree Regressor &   3.6443 &   316.4728 &  16.6898 &  0.5835 &  0.2528 &  0.2002 &     0.028 \\
omp      &      Orthogonal Matching Pursuit &  11.6651 &   661.4842 &  24.1136 &  0.4528 &  0.7900 &  1.4896 &     0.006 \\
lr       &                Linear Regression &  11.9426 &   801.5580 &  25.0735 &  0.3844 &  0.8129 &  1.7359 &     0.005 \\
ridge    &                 Ridge Regression &  11.9412 &   805.8845 &  25.1101 &  0.3820 &  0.8126 &  1.7356 &     0.006 \\
br       &                   Bayesian Ridge &  11.9338 &   845.8109 &  25.4381 &  0.3597 &  0.8106 &  1.7344 &     0.007 \\
huber    &                  Huber Regressor &   6.0315 &   928.7155 &  27.9140 &  0.3125 &  0.4323 &  0.4065 &     0.056 \\
par      &     Passive Aggressive Regressor &   9.6714 &  1020.8535 &  29.4623 &  0.2129 &  0.6885 &  0.9275 &     0.008 \\
llar     &     Lasso Least Angle Regression &  12.0764 &  1214.2287 &  32.9040 & -0.0066 &  0.9467 &  1.6193 &     0.006 \\
dummy    &                  Dummy Regressor &  12.0764 &  1214.2287 &  32.9040 & -0.0066 &  0.9467 &  1.6193 &     {\bf 0.003} \\
lasso    &                 Lasso Regression &  11.7063 &  1771.4781 &  30.7326 & -0.1609 &  0.7718 &  1.6734 &     0.009 \\
lar      &           Least Angle Regression &  16.1396 &  2069.0267 &  33.7691 & -0.4530 &  1.0109 &  2.5780 &     0.007 \\
en       &                      Elastic Net &  11.7823 &  3071.2942 &  36.0023 & -0.9102 &  0.7729 &  1.7297 &     0.008 \\
\hline
\end{tabular}
\end{table*}

The reference H$_2$ column density maps we use in this paper have an angular resolution of $18.2\arcsec$ and were derived from {\it Herschel} Gould Belt Survey (HGBS) data \citep{Andre10} and {\it Herschel} imaging survey of OB Young Stellar objects (HOBYS) data \citep{Motte10} by fitting the observed spectral energy distributions (SEDs) between 160 and 500 microns on a pixel-by-pixel basis and employing the multi-resolution procedure described in Appendix~A of \citet{Palmeirim13}. Since submillimeter dust continuum emission is optically thin and has a wide density dynamic range, it thus traces reliably the wide range of densities present in molecular clouds. The angular resolution of the H$_2$ column density is smoothed from the original angular resolution of 18$\arcsec$.2 to 25$\arcsec$ to be the same as that of the molecular line data.  The uncertainty of the H$_2$ column density derived from the HGBS data is a factor of 2 \citep{Roy14}.

The maps of $^{12}$CO, $^{13}$CO, and C$^{18}$O are made with a pixel size of 7$\arcsec$.5. The HGBS H$_2$ column density maps were regrid to align to the same grid as the molecular line data.

\section{Machine learning}\label{sec:machine_learning}

\subsection{Data preprocessing}

The images of the integrated intensity, peak intensity, ratio of $^{12}$CO to $^{13}$CO, the ratio of $^{13}$CO to C$^{18}$O, the excitation temperature derived using Eq. (\ref{EQ:Tex}), and H$_2$ column density shown in Fig. \ref{fig_orionA_obs} are converted into the data frame as listed in Table \ref{table:dataframe}. 
We made regressors trained on the data set for small portions of the OMC-1, OMC-1/2/3, NGC~2024, Aquila, and M~17 clouds, hereafter referred to as Regressor-OMC1, Regressor-OMC123, Regressor-NGC2024, Regressor-Aquila, and Regressor-M17, respectively. Table \ref{table:model_tuned_param} lists the trained area for each regressor. In Figures \ref{fig_OMC1_prediction}, \ref{fig_OMC123_prediction}, \ref{fig_ngc2024_prediction}, \ref{fig_aquila_prediction}, and \ref{fig_M17_prediction}, the area used for each regressor is indicated by the black box.

\begin{figure*}
\centering
\includegraphics[angle=0,width=16cm]{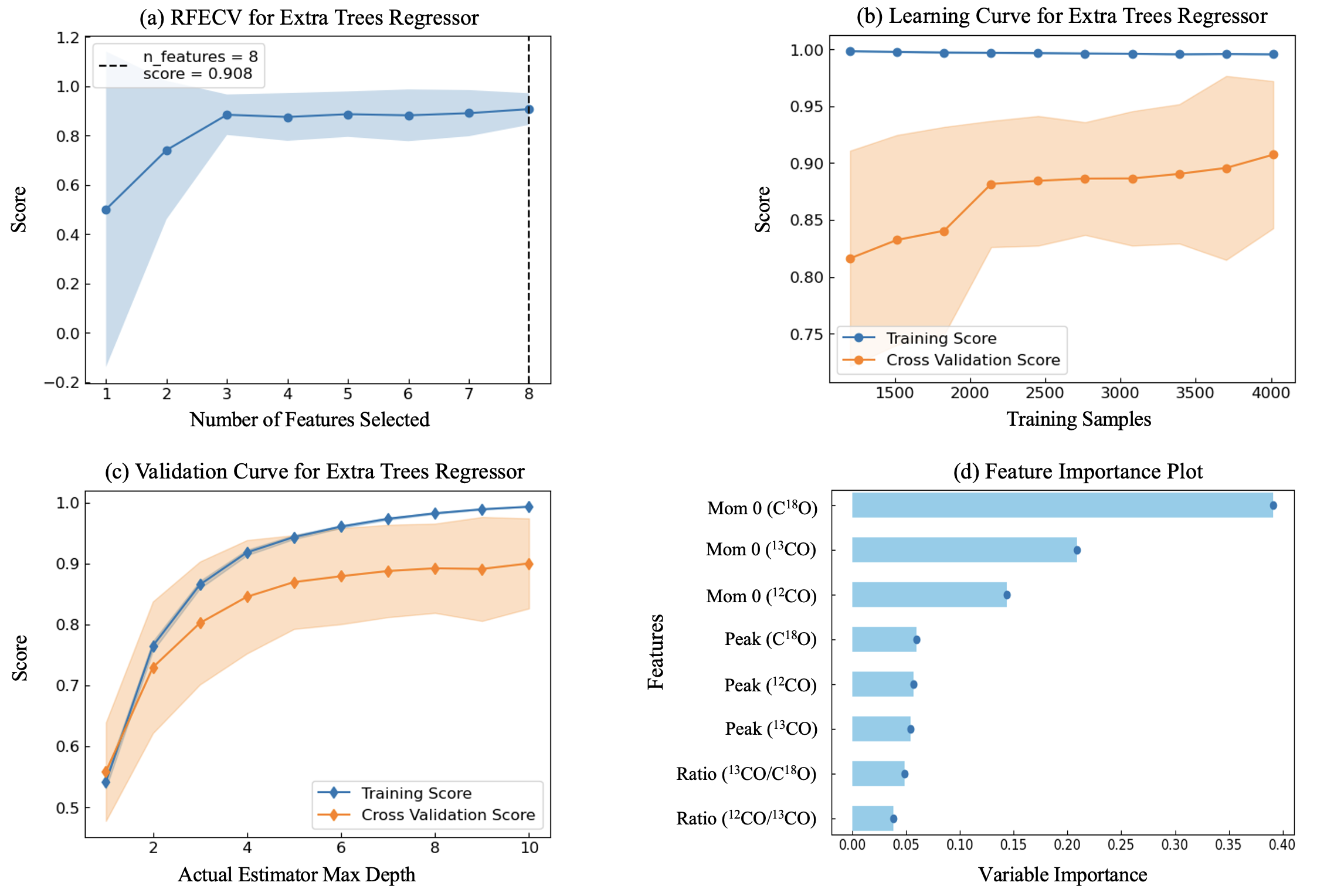}
\caption{Plots of ($a$) the recursive feature elimination with the cross-validation (RFECV), ($b$) learning curve, ($c$) validation curve, and ($d$) importance of each feature for Regressor-OMC1. The blue and orange curves indicate the training and cross-validation scores in panels ($b$) and ($c$). In panels ($a$), ($b$), and ($c$), the blue and orange filled curves indicate the range of the standard deviation of the score. The score indicates R2. 
}
\label{fig_validation}
\end{figure*}

\begin{figure*}
\centering
\includegraphics[angle=0,width=16cm]{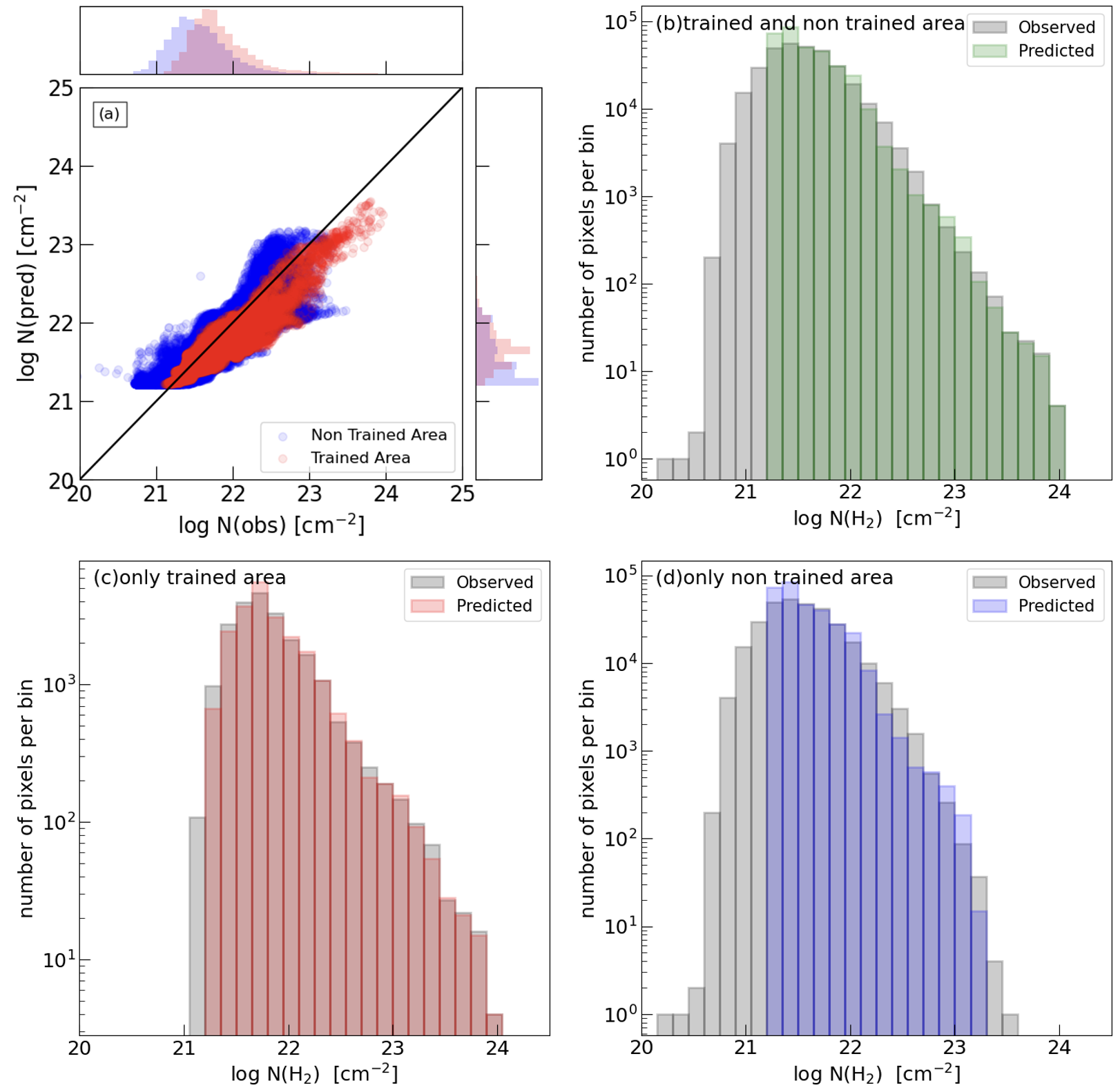}
\caption{($a$) Pixel-to-pixel correlation between predicted and observed H$_2$ column density and ($b$) pixel-by-pixel H$_2$ column density histogram for predicted and observed H$_2$ column density in the whole Orion~A region, ($c$) pixel-by-pixel H$_2$ column density histogram for predicted and observed H$_2$ column density in the area trained for Regressor-OMC1 (i.e., inside of the black box in Fig. \ref{fig_OMC1_prediction}$a$), and ($d$) pixel-by-pixel H$_2$ column density histogram for predicted and observed H$_2$ column density in the area not used for Regressor-OMC1 (i.e., outside of the black box in Fig. \ref{fig_OMC1_prediction}$a$). In panel ($a$), the black line indicates the predicted H$_2$ column density equals the observed H$_2$ column density. \textcolor{black}{The top and right on the panel ($a$) show the histograms of log $N$(obs) and log $N$(pred) for the non-trained area and trained area, respectively.} In panel ($b$-$d$), the red, green, and blue indicate the predicted H$_2$ column density, while the gray indicates the observed H$_2$ column density. 
}
\label{fig_OMC1_correlation_histgram}
\end{figure*}

\begin{figure*}
\centering
\includegraphics[angle=90,width=12.5cm]{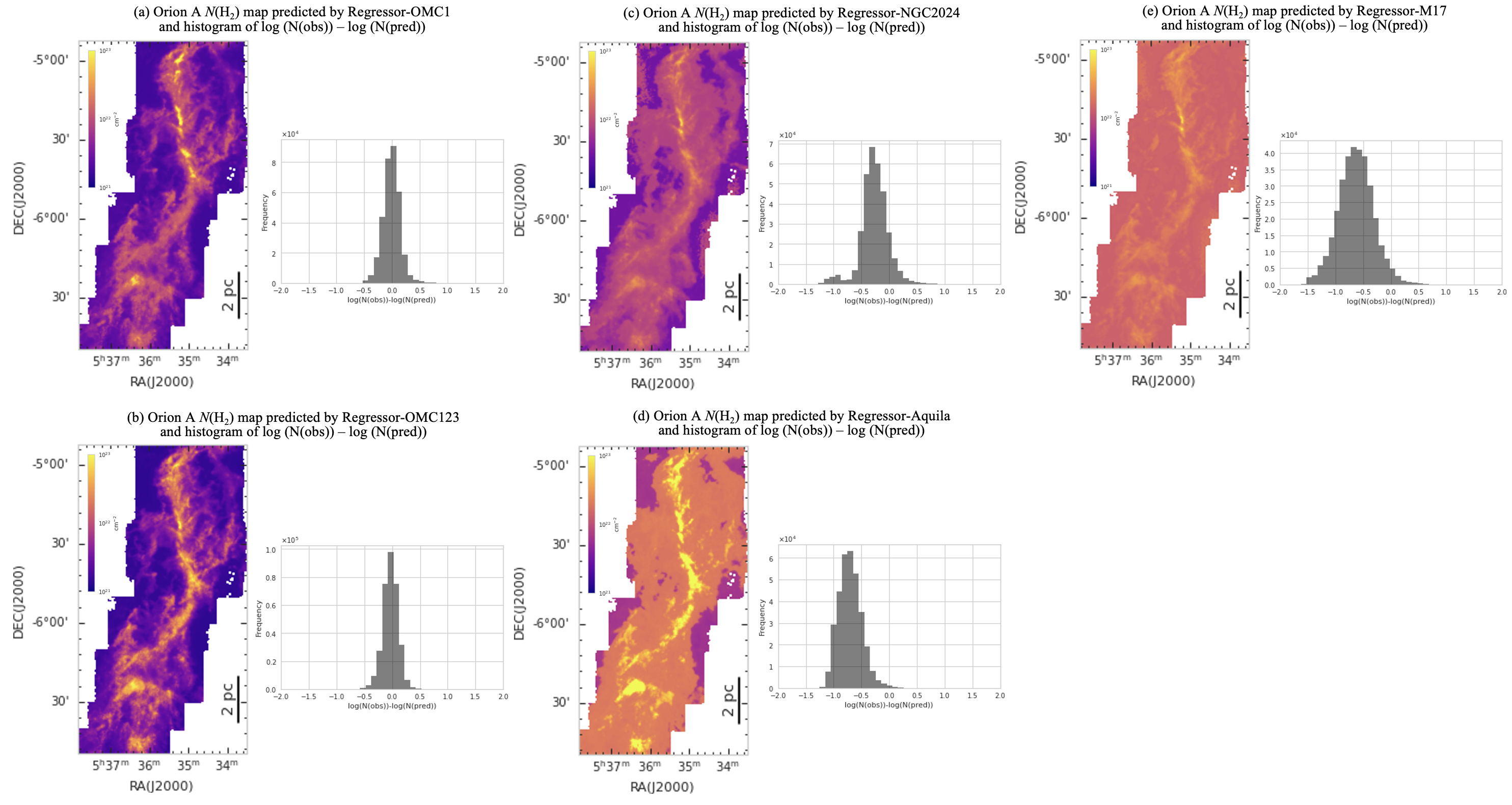}
\caption{Comparison of the predicted H$_2$ column density maps toward Orion~A predicted by using (a) Regressor-OMC1, (b) Regressor-OMC123, (c) Regressor-NGC2024, (d) Regressor-Aquila, and (e) Regressor-M17 \textcolor{black}{and histograms of log ($N$(obs)) $-$ log ($N$(pred))}.
}
\label{fig_results_among_model}
\end{figure*}

\subsection{Selection of the regression model}

In the python package {\it pycaret}, the available regression models are 
Extra Trees Regressor (ET), Random Forest Regressor (RF), Gradient Boosting Regressor (GBR), Decision Tree Regressor (DT), Light Gradient Boosting Machine (Lightt GMB), K Neighbors Regressor (KKN), AdaBoost Regressor (ADA), Lasso Regression (LASSO), Bayesian Ridge (BR), Linear Regression (LR), Ridge Regression (RIDGE), Elastic Net (EN), Orthogonal Matching Pursuit (OMP), Huber Regressor (HUBER), Least Angle Regression (LAR), Passive Aggressive Regressor (PAR), Lasso Least Angle Regression (LLAR), and Dummy Regressor (DUMMY).
To select the best model for this study, we evaluated the performance of these models using a function {\it compare\_models}. The {\it compare\_models} function trains and evaluates the performance of all available models using the $k$-hold cross-validation method  with the defined fold parameter (default = 10 folds, which was used in this study). The data are divided into $k$ subsets. One data subset is used for the test, while the $k$-1 subsets are used for training. The learning is continued $k$ times to make all $k$ data-set be used for testing.  Then, the accuracy averaged over $k$ training sessions is used for evaluation.
During the evaluation, the parameters are tuned to minimize the value of RMSLE (Root Mean Square Logarithmic Error). 
The ET was selected based on the comparison of RMSLE among models.
\textcolor{black}{The Extra Trees Regressor generates many decision trees, similar to the Random Forest Regressor \citep{Geurts2006}.
Random Forest is one of the most representative methods used in machine learning for classification, regression, and clustering.
It takes random samples from the entire data set, allowing for duplicates, and uses these samples to build multiple decision trees. 
Each tree is a hierarchical structure (flowchart-like tree structure) where an internal node represents a feature, the branch represents a decision rule, and each leaf node represents the outcome. Thus, each tree in the random forest is built on a subset sampled from the original data set and therefore has a different trend.
In the Extra Trees Regressor, a certain number of features within the entire set are randomly selected for each tree. The Extra Trees Regressor also randomly selects thresholds for each feature. These have the effect of avoiding overfitting.}
Table \ref{table:comp_model} shows RMSLE as well as MAE (Mean Absolute Error), MSE (Mean Square Error), RMSE (Root Mean Square Error), R2 (R-Squared), MAPE (Mean Absolute Percentage Error), and Training Time (TT) for each model. The definitions of MAE, MSE, RMSE, R2, RMSLE, and MAPE are given below.

\begin{equation}\label{eq:mae}
\centering
{\rm MAE} = \frac{1}{n} \sum_{i=1}^{n} | \hat{y}_i - y_i |,
\end{equation}

\begin{equation}\label{eq:MSE}
\centering
{\rm MSE} = \frac{1}{n} \sum_{i=1}^{n} (\hat{y}_i - y_i)^2,
\end{equation}

\begin{equation}\label{eq:RMSE}
\centering
{\rm RMSE} = \sqrt{\frac{1}{n} \sum_{i=1}^{n} (\hat{y}_i - y_i)^2},
\end{equation}

\begin{equation}\label{eq:R2}
\centering
{\rm R2} = 1- \frac{\sum_{i=1}^{n} (\hat{y}_i - y_i)^2}{\sum_{i=1}^{n} (\bar{y} - y_i)^2},
\end{equation}

\begin{equation}\label{eq:RMSLE}
\centering
\textcolor{black}{{\rm RMSLE} = \sqrt{\frac{1}{n} \sum_{i=1}^{n} (\log{\hat{y}_i - \log{y_i})^2}},}
\end{equation}

\begin{equation}\label{eq:MAPE}
\centering
{\rm MAPE} = \frac{1}{n} \sum_{i=1}^{n} (\frac{\hat{y}_i - y_i}{y_i})^2,
\end{equation}

\noindent where $n$, $\hat{y_i}$, $y_i$, $\bar{y}$ are the pixel number for each data, the predicted value for $i$th data, observed value for $i$th data, and the average of the observed values. These evaluation indexes, except R2, take lower values for better accuracy, while R2 ranges from 0 to 1 and takes values closer to 1 for better accuracy.
The model of Extra Trees Regressor (ET) shows the best scores in MAE, MSE, RMSE, RMSLE, and MAPE, except R2 as can be seen in Table \ref{table:comp_model}. We thus use ET in this study.

\subsection{Optimization of setup}

To tune the hyperparameters of the ET, we used a function {\it tune\_model} in the python package {\it pycaret}. This function tunes the hyperparameters automatically using the random grid search. 

In general, as increasing the number of features, the model becomes more complex and overfitted. Recursive feature elimination (RFE) is one of the methods to avoid overfitting. In this method, first, the model is produced using all features. Then, another model is produced using all features except the feature showing the lowest importance in the previous cycle. These steps are repeated. RFECV is the method in that recursive feature elimination (RFE) is processed with cross-validation (CV). 
Figure \ref{fig_validation} ($a$) shows the RFECV plot. The score increases as the number of selected features increases. As a result, eight features are selected for making the model. 
Figure \ref{fig_validation} ($b$) shows the learning curve. 
\textcolor{black}{The learning curve shows scores for different numbers of training samples.} 
At the beginning of the training, as updating the training model, the scores increase and the gap between the scores for the training data and test (cross-validation) data becomes small, suggesting the effect of the overfitting is improved. 
Figure \ref{fig_validation} ($c$) shows the validation curves when the hyperparameter {\it max\_depth} value changes\footnote{\textcolor{black}{The validation curve plot only displays numbers up to 1 less than the max\_depth value is due to the implementation of the function used to create the validation curve plot. In the case of max\_depth=11, the validation curve function tries max\_depth values from 1 to 10 and plots the cross-validation results.}}. This parameter plays a role in preventing overfitting. The score increases as increasing the number of max\_depth, while the gap between the scores of training and the test becomes large.  
Figure \ref{fig_validation} ($d$) shows the importance of each feature. The integrated intensity of C$^{18}$O plays a significant role in predicting the H$_2$ column density. The integrated intensity of $^{12}$CO and $^{13}$CO also plays a role. \textcolor{black}{In the case of Regressor-OMC1, the sum of the importance values for $^{12}$CO, $^{13}$CO, and C$^{18}$O integrated intensity maps is $\sim$0.75, while the sum for all feature is 1.0. Although the fraction of the importance values for other features is smaller than those for the integrated intensity maps, other features also play a role in the prediction. }
Table \ref{table:model_tuned_param} summarizes the parameters tuned for the given data set \footnote{\textcolor{black}{The best hyperparameters evaluate the performance of the model by trying combinations of hyperparameters. The model's performance is averaged over multiple training and evaluation runs using cross-validation.}}.

\begin{table*}  
\footnotesize
\caption{Fraction of predicted to observed total $N$(H$_2$) column density \\in the case of Regressor-OMC1}
\label{table:mass_estimate_OMC1}
\label{table:mass_estimate_OMC123}
\label{table:mass_estimate_NGC2024}
\label{table:mass_estimate_Aquila}
\label{table:mass_estimate_M17}
\centering  
\begin{tabular}{lcccccc}
\hline
    &   &  &  &  Regressor & & \\
{} &    & Regressor-OMC1 & Regressor-OMC123 & Regressor-NGC2024 & Regressor-Aquila & Regressor-M17 \\
\hline
\multirow{4}{*}{Predicted Area} & Orion~A &   97\%  &   109\% &   134\%  &   393\% &  278\% \\
                                & Orion~B/NGC~2024 &  133\% &  126\% & 95\% &  263\% &  266\% \\
                                & Aquila  &   80\% &   113\% &   86\%  &   102\% &   206\%  \\
                                & M~17     &  160\% &  153\% &  75\%   &  260\%  &  101\% \\
\hline
\end{tabular}
\end{table*}

\section{Results} \label{sec:results}

\subsection{Distribution of the predicted \ensuremath{\rm H_2} column density}
\subsubsection{Orion~A }\label{sec:results_orion}

Figure \ref{fig_OMC1_prediction} compares the predicted and observed H$_2$ column density maps of Orion~A. While only a small portion of the CO maps of the Orion A region, that with the highest H$_2$ column density (marked by a black rectangle), was used for training the regressor, the predicted H$_2$ column density map is quite similar to the reference H$_2$ column density map in most of the region.


Figure \ref{fig_OMC123_prediction} shows the results when the data around a larger area encompassing OMC-1/2/3 regions is used for the training. While the overall distribution of the predicted H$_2$ column density is similar to that of the observed H$_2$ column density, the prediction at higher H$_2$ column density is overestimated. 

Figure \ref{fig_OMC1_correlation_histgram}($a$) shows the pixel-to-pixel correlation between the observed H$_2$ column density and H$_2$ column density predicted by Regressor-OMC1. The predicted H$_2$ column density is mostly the same as the observed H$_2$ column density. For the observed H$_2$ column densities $< 2 \times 10^{21} {\rm cm}^{-2}$, the scatter of the predicted H$_2$ column densities is slightly larger than for the larger H$_2$ column density values because of the noise present in the molecular line data. 

Figure \ref{fig_OMC1_correlation_histgram}($b$) shows the pixel-by-pixel histogram of the observed H$_2$ column density and the H$_2$ column density predicted by Regressor-OMC1 toward the whole Orion~A region. The H$_2$ column density less than 10$^{21}$ cm$^{-2}$ is not predicted well by the regressor.

Figure \ref{fig_OMC1_correlation_histgram}($c$) shows the pixel-by-pixel histogram of the observed H$_2$ column density and the H$_2$ column density predicted by Model-OMC1 toward the area that is used for the training in Model-OMC1. The overall distribution between the observed and predicted H$_2$ column density is similar. 

Figure \ref{fig_OMC1_correlation_histgram}($d$) shows the pixel-by-pixel histogram of the observed H$_2$ column density and the H$_2$ column density predicted by Regressor-OMC1 toward the area that is not used for the training in Regressor-OMC1. The column densities of $<$ 10$^{21}$ cm$^{-2}$ and $>$ 5$\times$10$^{22}$ cm$^{-2}$ are not well predicted. These suggest that the prediction accuracy at the column densities of $<$ 10$^{21}$ cm$^{-2}$ and $>$ 5$\times$10$^{22}$ cm$^{-2}$ is low. \textcolor{black}{The reason why the column densities of $<$ 10$^{21}$ cm$^{-2}$ is not well predicted is that the training area does not contain low column density samples. The reason why the column densities of $>$ 5$\times$10$^{22}$ cm$^{-2}$ is not well predicted is discussed in Sect. \ref{sect5.2}. }


\textcolor{black}{Figure \ref{fig_results_among_model} ($a$) shows the predicted H$_2$ column density maps toward Orion A predicted by using Regressor-OMC1 and histograms of log ($N$(obs))- log ($N$(pred)). The residual of log ($N$(obs)) - log ($N$(pred)) shows the Gaussian distribution with a center of log ($N$(obs)) - log ($N$(pred)) = 0, suggesting the model makes a good prediction.}

\subsubsection{Orion~B/NGC~2024, Aquila, and M~17}

We also made regressors using a small portion of the Orion~B/NGC~2024, Aquila, and M~17 maps, respectively, and applied them to predict the H$_2$ column density toward the whole map. Figures \ref{fig_ngc2024_prediction}, \ref{fig_aquila_prediction}, and \ref{fig_M17_prediction} compare the predicted and observed H$_2$ column density maps toward Orion~B/NGC~2024, Aquila, and M~17. The overall distributions of the predicted H$_2$ column density in each region are similar to those of the observed H$_2$ column density. 
It is also worth mentioning that two condensations associated with the young stellar objects seen in the HGBS H$_2$ column density map are not seen in the predicted H$_2$ column density map of NGC~2024 (see magenta open circles in Fig. \ref{fig_ngc2024_prediction}). This implies that the conversion factor from the CO and its isotope's intensities to the H$_2$ column density are completely different (i.e., the relation between H$_2$ column density and molecular intensities around these two condensations is different compared with the overall environment). This also may suggest that the prediction of the H$_2$ column density by machine learning is helpful in finding regions of the cloud with atypical gas properties.

\begin{figure}
\centering
\includegraphics[angle=0,width=9cm]{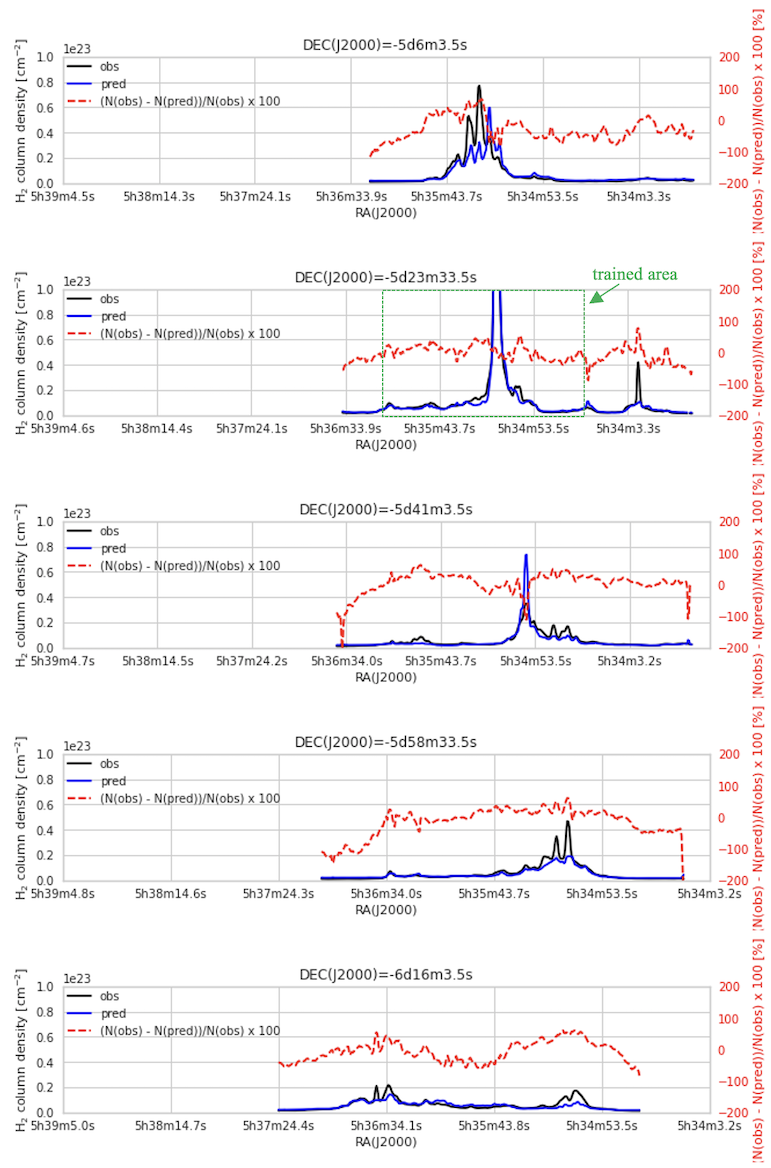}
\caption{Profiles of the H$_2$ column density along the RA direction. The Dec of cut lines is indicated at the top of each panel.
The black and blue curves indicate the profiles of the observed and predicted H$_2$ column density ($N{\rm (obs)}$ and $N{\rm (pred)}$). 
The red curve indicate fraction of $N{\rm (obs)}-N{\rm (pred)}/N{\rm (obs)} \times 100$. The green box indicates the area that is used for training.
}
\label{fig_OMC1_profile}
\end{figure}

\section{Discussion}\label{sect:discuss}

\subsection{Comparison between the observed and predicted \ensuremath{\rm H_2} column density maps}

\textcolor{black}{Table \ref{table:mass_estimate_OMC1} summarizes} the predicted total H$_2$ column density fraction to the observed total H$_2$ column density. 
\textcolor{black}{The predicted total column density, $N(\rm pred)_{tot}$, is derived using the equation $N(\rm pred)_{tot}$=$\sum N(\rm pred)$, where $N(\rm pred)$ is the predicted column density at each pixel. 
The observed total column density, $N(\rm obs)_{tot}$, is derived using the equation $N(\rm obs)_{tot}$=$\sum N(\rm obs)$, where $N(\rm obs)$ is the observed column density at each pixel. The mass of each cloud, $M_{\rm cloud}$ is estimated as
\begin{equation}
M_{\rm cloud} = A N_{\rm tot} m_{\rm H} \mu_{\rm H_2},
\end{equation}
,where $A$ is the surface area, $m_{\rm H}$ is the hydrogen atom mass, and $\mu_{\rm H_2}=2.8$ is the mean molecular weight per H$_2$ molecule \citep{Shimajiri17}. Thus, on the comparison toward the same area, the relation between $N(\rm pred)_{tot}$ and $N(\rm obs)_{tot}$ is proportional to the relation between the predicted mass and observed mass.
}
If the area used for the training is inside the area where the H$_2$ column density is predicted, the fraction ranges from 95\% to 109\%. 
However, if the area used for the training is not inside the area where the H$_2$ column density is predicted, significant variations in the fraction are seen. This indicates that variations in the molecular abundance among clouds are not corrected. However, the H$_2$ column density in Orion~B/NGC~2024 predicted by Regressor-NGC2024, Regressor-OMC1, and Regressor-OMC123 is relatively similar (95\%, 133\%, and 126\%), suggesting similar gas properties/abundance in these clouds. 

This analysis suggests that H$_2$ column densities can be well reconstructed using machine learning when the training data sets have similar properties as those of the clouds for which the H$_2$ column density is reconstructed. In the case where the training sets and the studied clouds have different gas properties, the machine learning method is not optimum. This can be improved by increasing the number of clouds with different properties in the training step and the molecular species used for the training. 

Here, we also compared the fraction of the predicted total H$_2$ column density to the observed total H$_2$ column density ($N(\rm pred)_{\rm tot}$/$N(\rm obs)_{\rm tot}$) with that of the total H$_2$ column density derived from $^{12}$CO (1--0) integrated intensity using $X_{\rm CO}$ factor to the observed total H$_2$ column density ($N(X_{\rm CO})_{\rm tot}$/$N({\rm obs})_{\rm tot}$). As summarized in Table \ref{table:mass_comp}, the fraction of $N(X_{\rm CO})_{\rm tot}$/$N(\rm obs)_{\rm tot}$ varies from 70\% to 257\%, while the fraction of $N(\rm pred)_{\rm tot}$/$N(\rm obs)_{\rm tot}$ varies from 95\% to 110\%. This comparison suggests that the Machine learning technique provides the total cloud mass with a smaller uncertainty than the total cloud mass estimate using $X_{\rm CO}$.

\subsection{Comparison of the observed and predicted \ensuremath{\rm H_2} column density profiles across the maps}\label{sect5.2}

As described in Sect. \ref{sec:results_orion}, there is a trend that the accuracy of predicting H$_2$ column density is lower at higher H$_2$ column density. To investigate which portion of the could structures is lower accuracy in the predicted H$_2$ column density, we produced profiles along the RA direction as shown in Fig. \ref{fig_OMC1_profile}. 
The distributions of the profile at H$_2$ column density $N(\rm H_2) < \sim 1 \times 10^{22}$ cm$^{-2}$ are similar between the predicted and observed H$_2$ column density, while the peaks with H$_2$ column density $N(\rm H_2) > \sim 2 \times 10^{22}$ cm$^{-2}$ is not predicted well. These peaks correspond to the dense star-forming gas often organized in filamentary structures. The reason why the higher H$_2$ column density is not predicted well is that the $^{12}$CO, $^{13}$CO, and C$^{18}$O used in training do not trace well the dense gas \citep{Shimajiri15a}. Thus, to predict the higher H$_2$ column density, dense gas tracers such as H$^{13}$CO$^+$ and N$_2$H$^+$ are required. \citet{Shimajiri17} reported that the H$^{13}$CO$^+$ (1--0) emission traces HGBS filaments very well above $A_{\rm V}$ $>$ 16.
The results by \citet{Gratier21} support our results (also see Sect. \ref{sect:introduction}).

\begin{table*}  
\caption{Comparison of predicted, observed, and $X_{\rm CO}$-derived total H$_2$ column densities across entire fields}
\label{table:mass_comp}
\centering  
\begin{tabular}{lrrrrr}
\hline
{Predicted Area}                    &  Orion~A &  Orion~A &  NGC~2024 &  Aquila &     M~17 \\
\hline
{Regressor}                      &  Regressor-OMC1 & Regressor-OMC123 &  Regressor-NGC2024 &  Regressor-Aquila &     Regressor-M17 \\
\hline
$N(\rm pred)_{\rm tot}$$^{\dag}$/$N(\rm obs)_{tot}$$^{\ddag}$          &   97\% &   109\% &     95\% &     102\% &    101\% \\
$N(X_{\rm CO})_{\rm tot}$$^{\spadesuit}$/$N(\rm obs)_{tot}$$^{\ddag}$     &  257
\% &  257\% &   394\% &     70\% &  222\% \\
\hline
\multicolumn{6}{l}{$^{\dag}$:$N(\rm pred)_{tot}$ is the predicted total column density. $N(\rm pred)_{tot}$=$\sum N(\rm pred)$, where $N(\rm pred)$ is the predicted column density at each pixel.} \\
\multicolumn{6}{l}{$^{\ddag}$:$N(\rm obs)_{tot}$ is the observed total column density. $N(\rm obs)_{tot}$=$\sum N(\rm obs)$, where $N(\rm obs)$ is the observed column density at each pixel.}\\
\multicolumn{6}{l}{$^{\spadesuit}$:$N(X_{\rm CO})_{\rm tot}$ is the $X_{\rm CO}$-derived total column density. $N(X_{\rm CO})_{\rm tot}$=$\sum N(X_{\rm CO})$, where $N(X_{\rm CO})$ is the $X_{\rm CO}$-derived column density at each pixel.}\\
\end{tabular}
\end{table*}

\subsection{Application of trained regressors to other clouds}

To investigate whether we can predict the H$_2$ column density for different clouds from those in which they were trained, we applied each model to the data set of all other clouds.
As shown in Fig. \ref{fig_results_among_model}, the distributions of the predicted H$_2$ column density of each cloud are not the same when different training sets are used. The fraction of the H$_2$ column density predicted by each model to the observed H$_2$ column density in each cloud is listed in \textcolor{black}{Table \ref{table:mass_estimate_OMC1}}. 
When the Regressor-OMC1 is applied to the data set of Orion~B/NGC~2024, Aquila, and M~17, the fractions of the predicted H$_2$ column density to the observed H$_2$ column density are 133\%, 80\%, and 160\%, respectively (\textcolor{black}{column Regressor-OMC1 in Table \ref{table:mass_estimate_OMC1}}). 
When the Regressor-OMC123 is applied to the data set of Orion~B/NGC~2024, Aquila, and M~17, the fractions are 126\%, 113\%, and 153\%, respectively (\textcolor{black}{column Regressor-OMC123 in Table \ref{table:mass_estimate_OMC123}}).
When the Regressor-NGC2024 is applied to the data set of Orion~A, Aquila, and M~17, the fractions are 134\%, 86\%, and 75\%, respectively (\textcolor{black}{column Regressor-NGC2024 in Table \ref{table:mass_estimate_NGC2024}}).
When the Regressor-Aquila is applied to the data set of Orion~A, Orion~B/NGC~2024, and M~17, the fractions of the predicted H$_2$ column density to the observed H$_2$ column density are 393\%, 263\%, and 260\%, respectively (\textcolor{black}{column Regressor-Aquila in Table \ref{table:mass_estimate_Aquila}}).
One of the problems of Aquila is that the background in the dust continuum is very strong \citep{Konyves15}, thus not all the column density derived from {\it Herschel} comes from the cloud emission but is rather a result of the line of sight integration. That may be a reason for this overestimation.
When the Regressor-M~17 is applied to the data set of Orion~A, Orion~B/NGC~2024, and Aquila, the fractions are 278\%, 266\%, and 206\%, respectively (\textcolor{black}{column Regressor-M17 in Table \ref{table:mass_estimate_M17}}). 
The large variations in the fraction (75--393\%) are recognized. However, the fractions in the cases where the regressor trained in Orion~A (OMC-1 and OMC-123) applied to Orion~B/NGC~2024 and the regressor trained in Orion~B/NGC~2024 applied to Orion~A tend to be smaller (95-134\%). These results indicate that the abundances of the molecule used in this study vary among the clouds and the current method using machine learning could not correct these variations. Orion~A and Orion~B are in the same giant molecular cloud complex. Thus, similar abundances in Orion~A and Orion~B are expected compared with those in other clouds. 
To test this hypothesis, we made a regressor trained in Orion~A, Aquila, and M~17, then applied the model to NGC~2024. But, the fraction and distribution of the predicted H$_2$ column density were not improved. Training in more clouds might be required to correct the abundance variations among clouds and also using more molecular species tracing different densities and different environments.

\begin{figure*}
\centering
\includegraphics[angle=0,width=16cm]{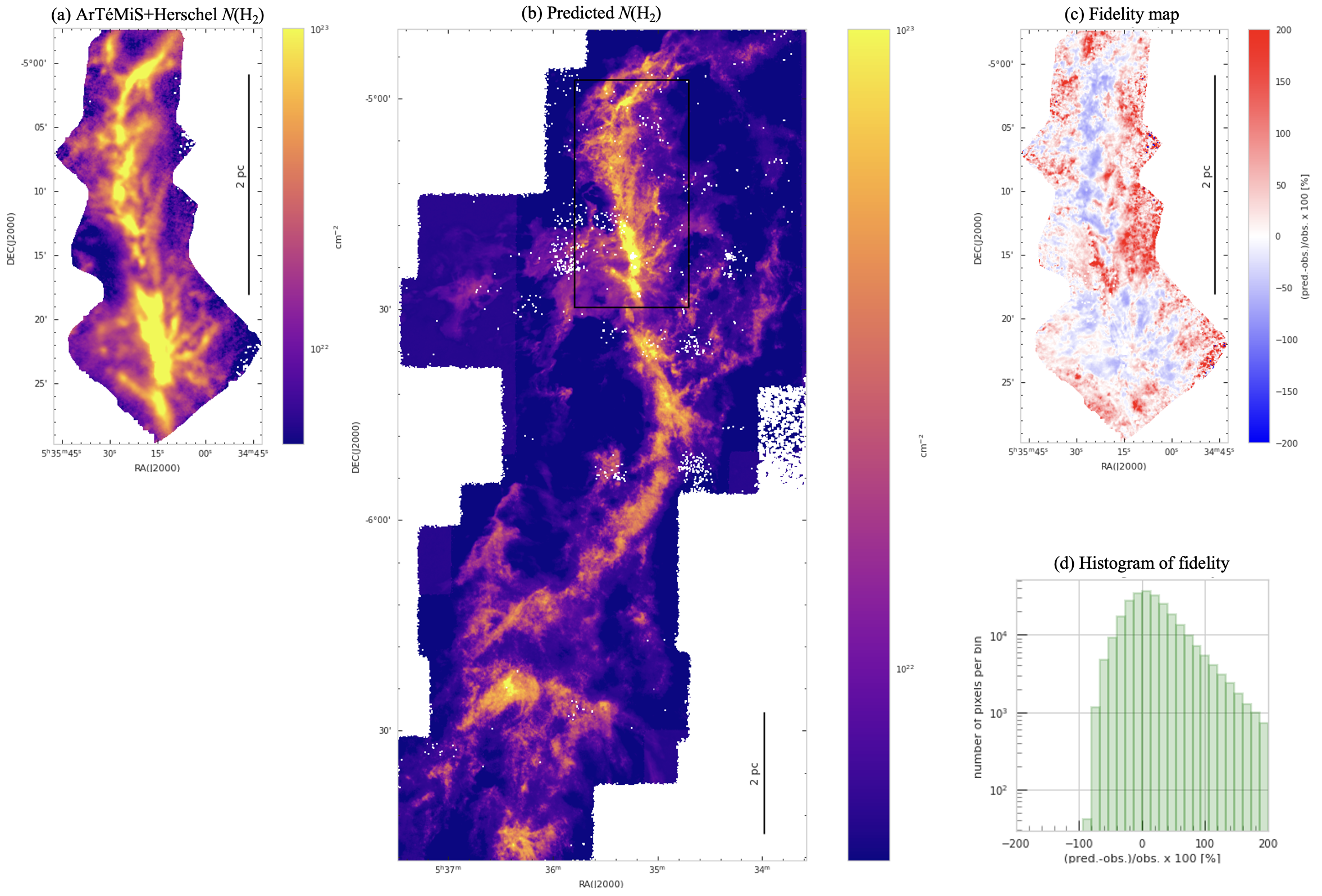}
\caption{Same as Figure \ref{fig_OMC1_prediction}, but predicted by Regressor-High-Resolution. The displayed area in panels ($a$) and ($c$) corresponds to the area indicated by the black box in panel ($b$).
}
\label{fig_carma_prediction}
\end{figure*}

\begin{figure*}
\centering
\includegraphics[angle=0,width=16cm]{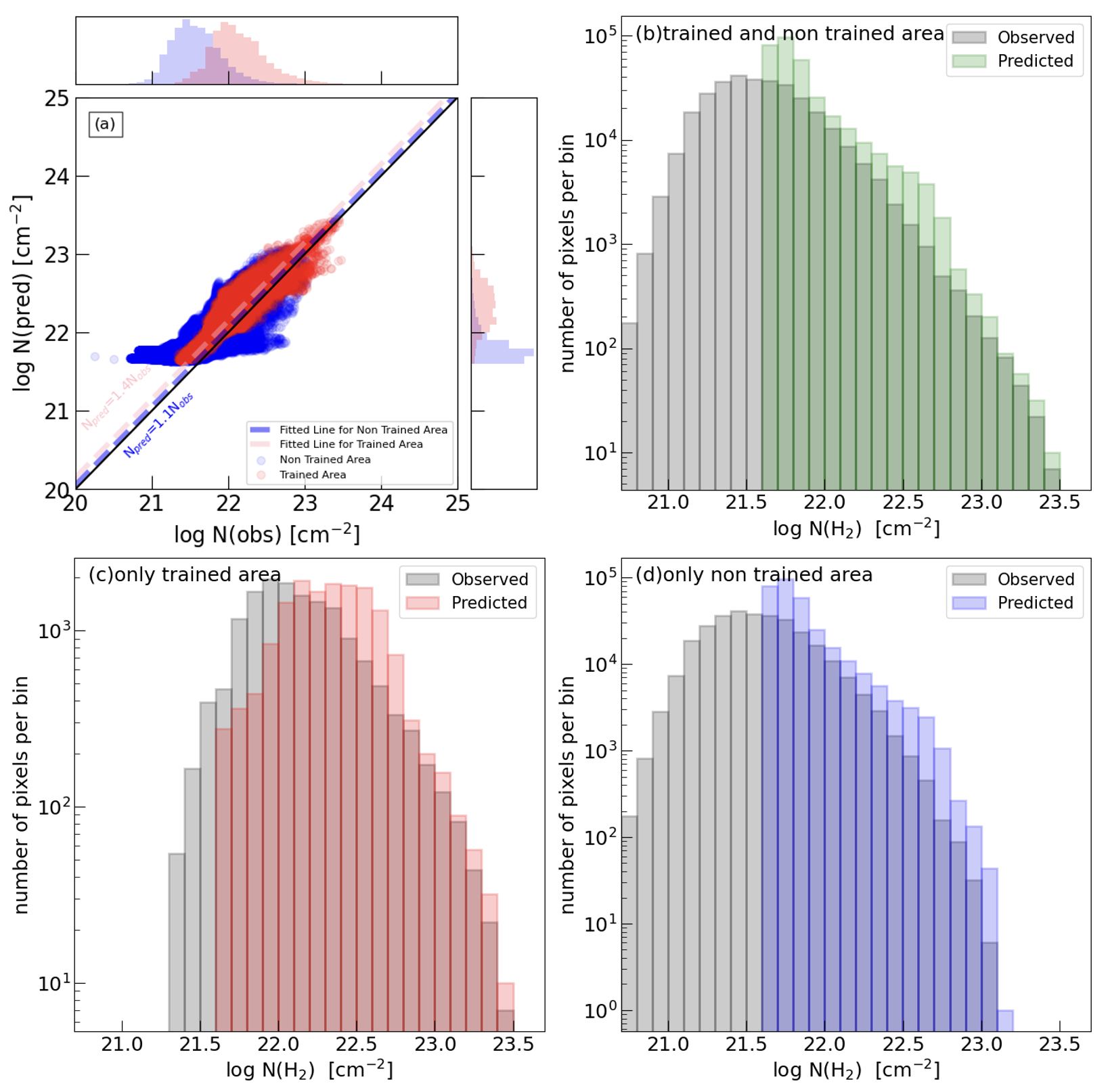}
\caption{($a$) Pixel-to-pixel correlation between predicted and observed H$_2$ column density on the 18"2 resolution and ($b$) pixel-by-pixel H$_2$ column density histogram for predicted and observed H$_2$ column density in the whole Orion~A region, ($c$) pixel-by-pixel H$_2$ column density histogram for predicted and observed H$_2$ column density in the area trained for Regressor-High-Resolution (i.e., inside of the black box in Fig. \ref{fig_carma_prediction} (a)), and ($d$) pixel-by-pixel H$_2$ column density histogram for predicted and observed H$_2$ column density in the area not used for Regressor-High-Resolution (i.e., outside of the black box in Fig. \ref{fig_carma_prediction} (a)). In panel ($a$), the black line indicates the predicted H$_2$ column density equals the observed H$_2$ column density. 
In panel ($a$), the red and blue dashed lines indicate the best fits results for the trained area and non-trained area: $N_{\rm pred}=1.4 \times N_{\rm obs}$ and $N_{\rm pred}=1.1 \times N_{\rm obs}$. \textcolor{black}{The top and right on the panel ($a$) show the histograms of log $N$(obs) and log $N$(pred) for the non-trained area and trained area, respectively.} 
In panel ($b$-$d$), the red, green, and blue indicate the predicted H$_2$ column density, while the gray indicates the observed H$_2$ column density.
}
\label{fig_CARMA_correlation_histgram}
\end{figure*}

\subsection{Predicting $N$(H$_2$) map with higher-angular resolution}

In Section \ref{sec:results}, we trained and created regressors by adjusting H$_2$ column densities in a small portion of each cloud where both molecular line data and {\it Herschel} data were available. Then, a H$_2$ column density map for an area much wider than that used for training was predicted from the molecular line data by applying the trained regressor. This indicates that our method can successfully predict the H$_2$ column density in an area not observed by {\it Herschel} and can also generate H$_2$ column density maps at higher angular resolution if {\it Herschel} column density data are available for a small portion of a given field 
and molecular line data exist for a larger area. 

\citet{Schuller21} observed the northern part of the Orion~A molecular cloud, which covers the OMC-1, OMC-2, and OMC-3 regions with the ArT$\acute{\rm e}$MiS\footnote{See \url{https://www.apex-telescope.org/ns/artemis/}\\ArT$\acute{\rm e}$MiS stands for “ARchitectures de bolom\`{e}tres pour des TElescopes\`{a} grand champ de vue dans le domaine sub-MIllim$\acute{\rm e}$trique au Sol” in French.} camera 
(see \citealp{Andre16}) 
installed on the Atacama Pathfinder EXperiment (APEX) telescope 
and produced a H$_2$ column density map of the integral-shaped filament 
at 8$\arcsec$ resolution. \citet{Kong08} obtained  $^{12}$CO (1--0), 
$^{13}$CO (1--0), and C$^{18}$O (1--0) maps of the Orion~A cloud at 8$\arcsec$ resolution, which cover a larger area than the ArT$\acute{\rm e}$MiS camera map,  
by combining $^{12}$CO (1--0), $^{13}$CO (1--0), and C$^{18}$O (1--0) data 
from the Nobeyama 45m telescope with interferometric data from the Combined Array for Research in Millimeter-wave Astronomy (CARMA).

First, we generated a $N$(H$_2$) predictor (Regressor-High-Resolution) by training the regressor over the entire area mapped with ArT$\acute{\rm e}$MiS. We then used this regressor to predict H$_2$ column density values in areas observed 
with CARMA but not covered by ArT$\acute{\rm e}$MiS, resulting in a wide-field,   8$\arcsec$ resolution $N$(H$_2$) map. 
Figure \ref{fig_carma_prediction} shows the ArT$\acute{\rm e}$MiS+{\it Herschel} 8$\arcsec$ resolution $N$(H$_2$), the predicted  8$\arcsec$ resolution $N$(H$_2$), the fidelity maps, and the fidelity histogram. The pixels covering the lower column density are limited on the 
ArT$\acute{\rm e}$MiS+{\it Herschel} 8$\arcsec$ resolution $N$(H$_2$) map. In other words, the training is insufficient at low column densities. 
This causes an overestimation of the predicted H$_2$ column density at low column densities (see Figure~\ref{fig_carma_prediction} (d)). However, the overall distribution of the integral-shape filament is well predicted 
(Figure~\ref{fig_carma_prediction} (c)). 

To evaluate whether the $8\arcsec$ angular resolution $N(\rm H_2)$ map is predicted well, 
we compared the predicted H$_2$ column density map smoothed to 18$\arcsec$.2 
from 8$\arcsec$ with the 18$\arcsec$.2 resolution {\it Herschel} H$_2$ 
column density map. Figure \ref{fig_CARMA_correlation_histgram} shows the 
pixel-to-pixel correlation between the predicted and observed H$_2$ column densities on the 18$\arcsec$.2 resolution and their histograms. The overall distribution is in good agreement, although the predicted H$_2$ column density in the trained and non-trained areas tends to be 10\% and 40\% larger than the observed H$_2$ column density, respectively
(see Fig. \ref{fig_CARMA_correlation_histgram}(a)). 
In particular, for $N$(H$_2$) $<$ 2$\times$10$^{21}$ cm$^{-2}$, the predicted H$_2$ column density is almost an order of magnitude larger than the observed H$_2$ column density. The reason is that the area of $N$(H$_2$) $<$ 2$\times$10$^{21}$ cm$^{-2}$ 
is not trained (see red points in Fig. \ref{fig_CARMA_correlation_histgram}(a)), 
and the Regressor-High-Resolution cannot accurately predict the 
H$_2$ column density of $N$(H$_2$) $<$ 2$\times$10$^{21}$ cm$^{-2}$.

As discussed above, the machine-learning predictions of H$_2$ column density made in this study allow us to do the following: 
(1) extend the coverage of the H$_2$ column density maps obtained with 
the {\it Herschel} space observatory by using ground-based data, 
and (2) enable the production of H$_2$ column density maps with an angular resolution higher than 18$\arcsec$.2, which could not be obtained only with the {\it Herschel} telescope, from the molecular line data obtained with interferometers such as CARMA and ALMA\footnote{ALMA stands for the Atacama Large Millimeter/submillimeter Array}.

\section{Conclusions}\label{sect:concle}

To investigate the potential of machine learning in producing reliable H$_2$ column density maps, we generated a number of H$_2$ column density maps from $^{12}$CO (1--0), $^{13}$CO (1--0), and C$^{18}$O (1--0) data using the python package {\it pycaret}. Our main results can be summarized as follows:

\begin{enumerate}

\item For the given data set, the model of the Extra Trees Regressor (ET) showed the best score in MAE, MSE, RMSE, RMSLE, and MAPE, although the training time tends to be longer than other models. In this study, we used the ET model. 

\item The overall distribution of the predicted H$_2$ column density is similar to that of {\it Herschel}-derived column density maps. The predicted total H$_2$ column density is also consistent within 10\% compared with the observed total H$_2$ column density when we created the model by training the small portion of the trained area in the same cloud. The ratio of the predicted H$_2$ column density to the observed H$_2$ column density ranges from 95\% to 110\%, while the ratio of the H$_2$ column density derived using the $X_{\rm CO}$ factor to the observed H$_2$ column density ranges from 70\% to 257\%. 

\item Comparing the predicted and observed H$_2$ column density maps in NGC~2024, we found that two condensations associated with the young stellar objects seen in the HGBS H$_2$ column density map are not seen in the predicted H$_2$ column density map. This implies that the conversion
factor from the CO and its isotope’s intensities to the H$_2$ column density are completely different (i.e., the relation between H$_2$ column density and molecular intensities is different). This also may suggest that the prediction of the H$_2$ column density by machine learning is helpful in finding such unique regions in the clouds.

\item We also investigated whether we can predict the H$_2$ column density for clouds different from those in which they were trained. Do to so, we applied each data set model to all the other clouds. 
We found that the H$_2$ column density is not predicted well by using the model trained in the data set of another cloud, suggesting that the abundances of the molecule used in this study vary among the clouds and the machine learning could not correct these variations.

\item We succeed in producing an 8$\arcsec$ resolution $N$(H$_2$) map from the $^{12}$CO, $^{13}$CO, and C$^{18}$O (1--0) line data obtained with CARMA and the Nobeyama 45m telescope. This indicates that the machine-learning technique in this study allows us to extend the H$_2$ column densities map obtained with the {\it Herschel} Space Observatory by using ground-based radio telescopes and to produce a H$_2$ column density map with an angular resolution higher than 18$\arcsec$.2, which could not be obtained only with the {\it Herschel} telescope, from the molecular line data obtained with interferometers.

\end{enumerate}

\section*{Acknowledgements}
\textcolor{black}{We thank the referee for useful suggestions that improved the clarity of the paper.}
This work was supported by JSPS KAKENHI Grant Numbers JP19K23463, JP20K04035, and JP21H00057. This work was also supported by “Young interdisciplinary collaboration project” in the National Institutes of Natural Sciences (NINS). This work was supported (in part) by a University Research Support Grant from the National Astronomical Observatory of Japan (NAOJ). This research has made use of data from the HGBS and HOBYS projects.
The authors are grateful to Kaoru NISHIKAWA and Daisuke YOSHIDA for the useful discussion.

\section*{DATA AVAILABILITY}
The data underlying this article are available in the article.



\bibliographystyle{mnras}
\bibliography{Shimajiri.bbl} 



\appendix

\section{Some extra material}

\begin{figure*}
\centering
\includegraphics[angle=0,width=14cm]{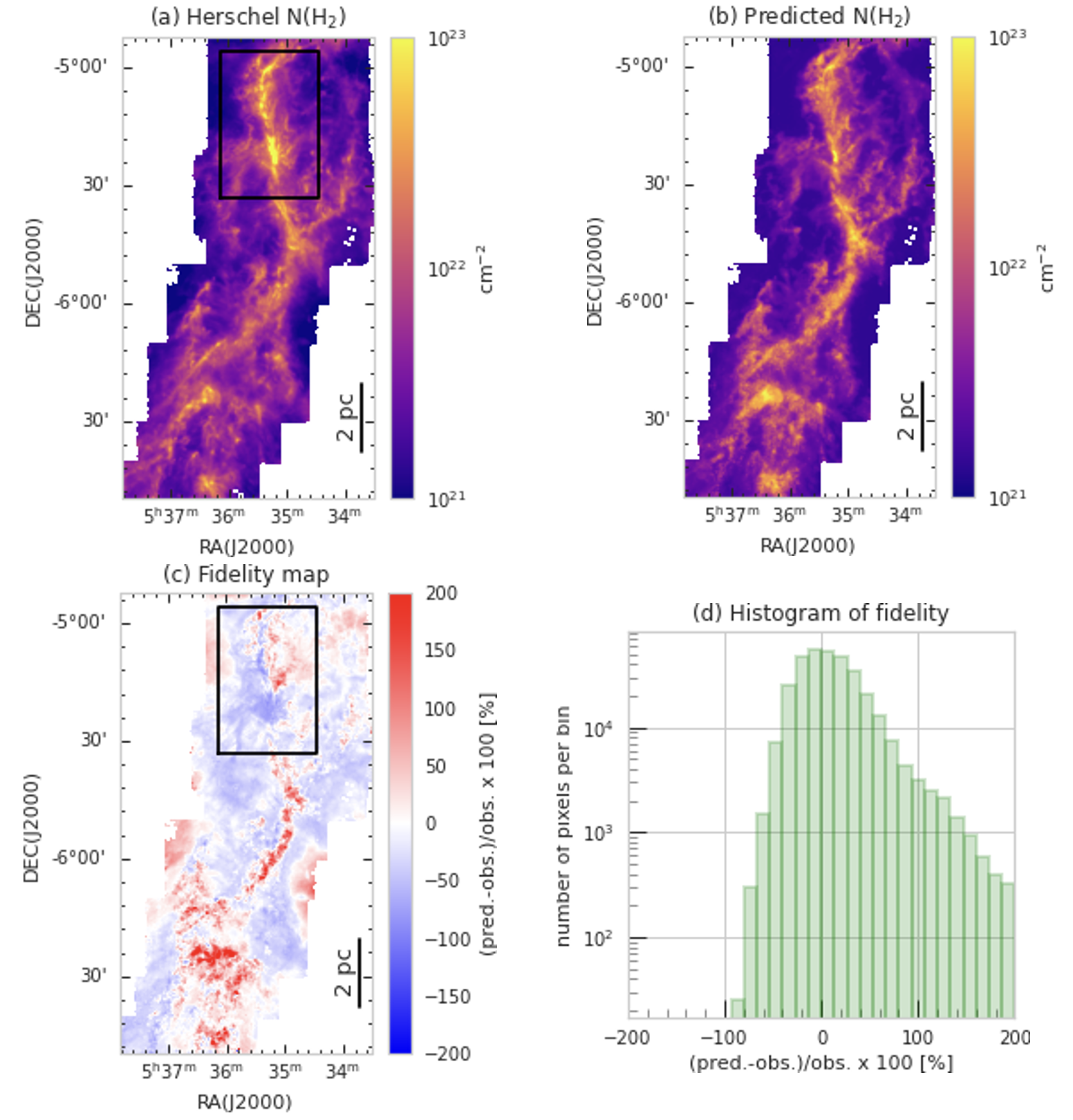}
\caption{Same as Figure \ref{fig_OMC1_prediction}, but predicted by Regressor-OMC123.
}
\label{fig_OMC123_prediction}
\end{figure*}

\begin{figure*}
\centering
\includegraphics[angle=0,width=14cm]{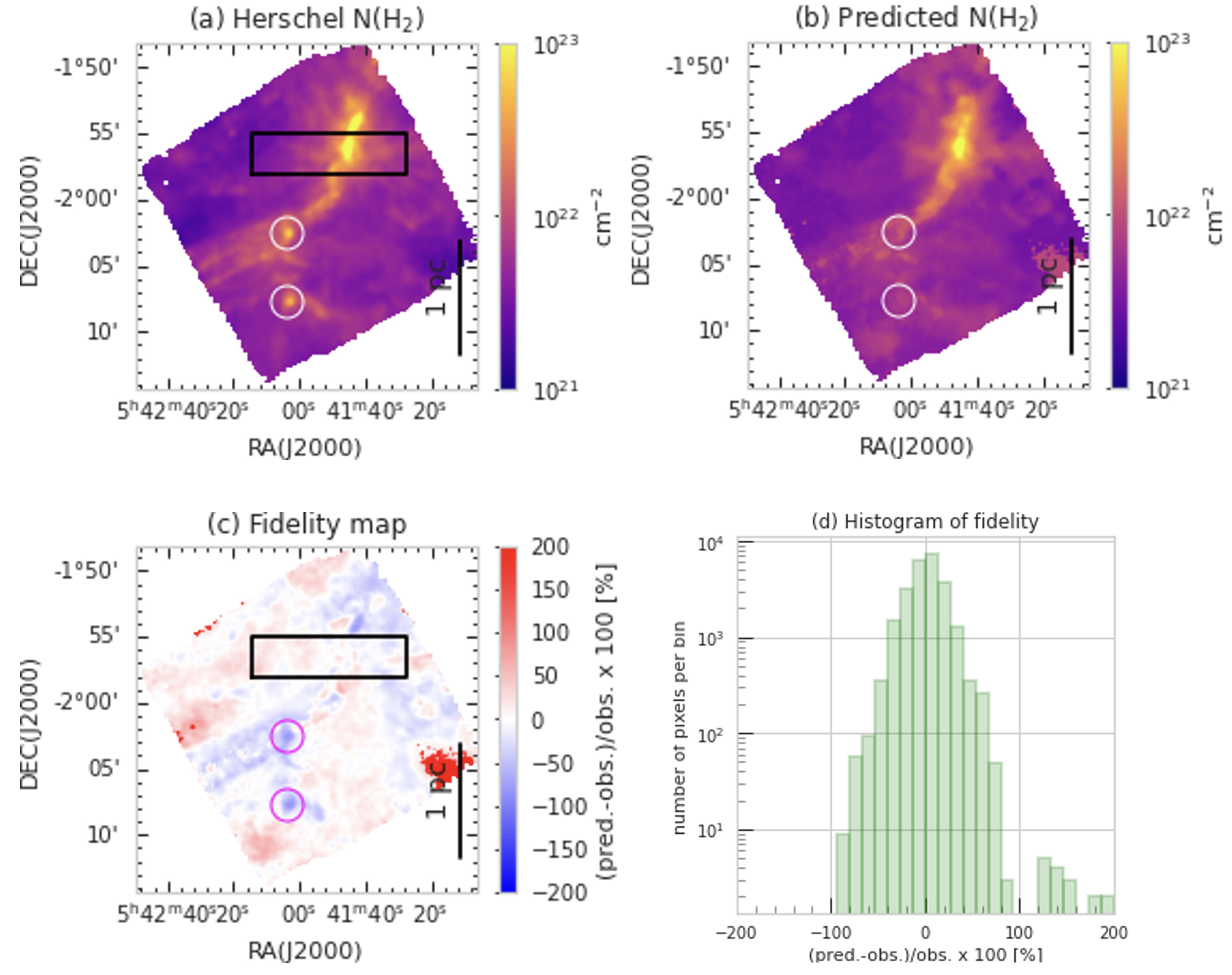}
\caption{Same as Figure \ref{fig_OMC1_prediction}, but toward Orion~B/NGC~2024 and predicted by Regressor-NGC2024. The \textcolor{black}{white} open circles indicate the location of two condensations seen in the observed H$_2$ column density map and not in the predicted H$_2$ column density map.
}
\label{fig_ngc2024_prediction}
\end{figure*}

\begin{figure*}
\centering
\includegraphics[angle=0,width=17cm]{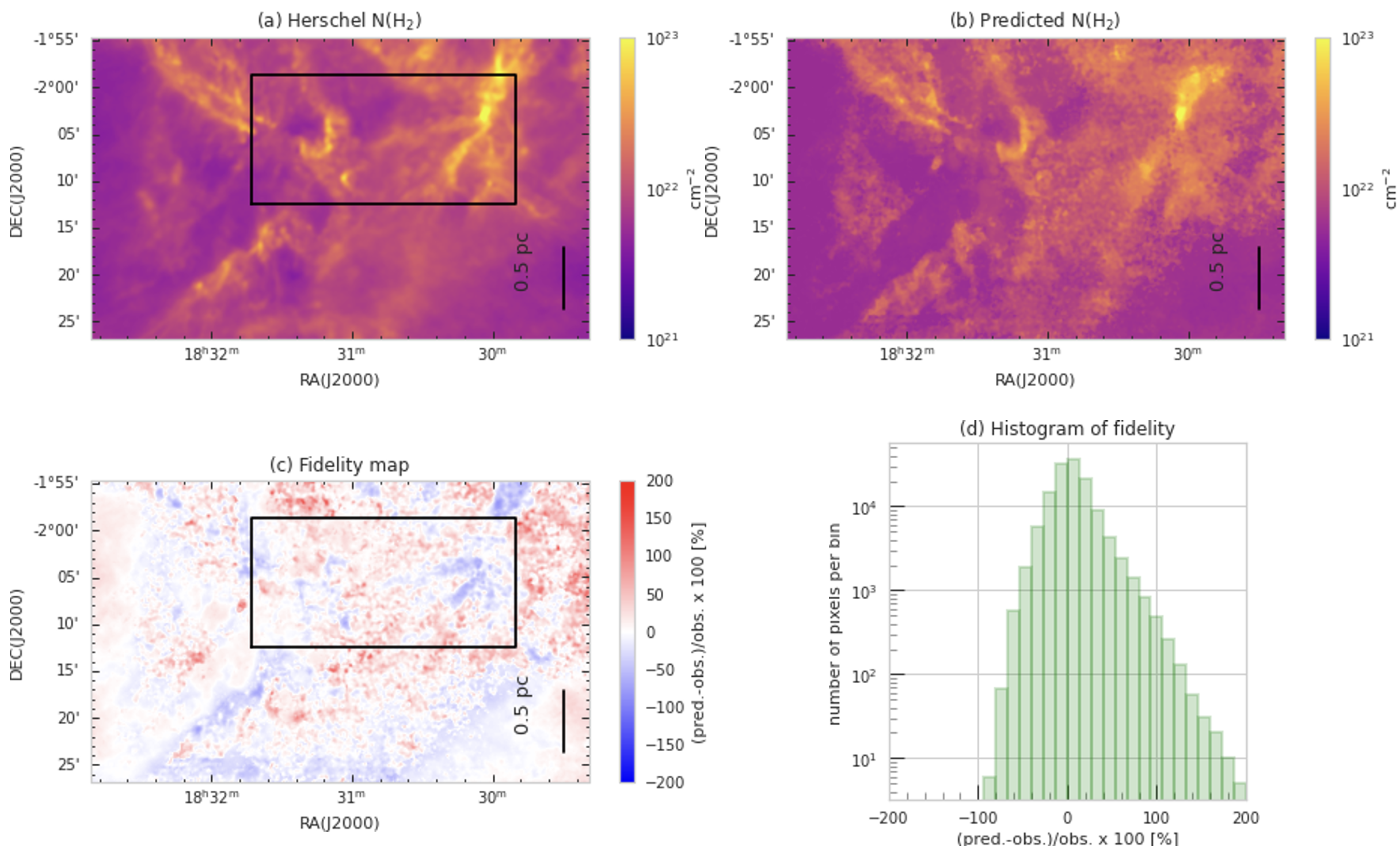}
\caption{Same as Figure \ref{fig_OMC1_prediction}, but toward Aquila and predicted by Regressor-Aquila.
}
\label{fig_aquila_prediction}
\end{figure*}

\begin{figure*}
\centering
\includegraphics[angle=0,width=18cm]{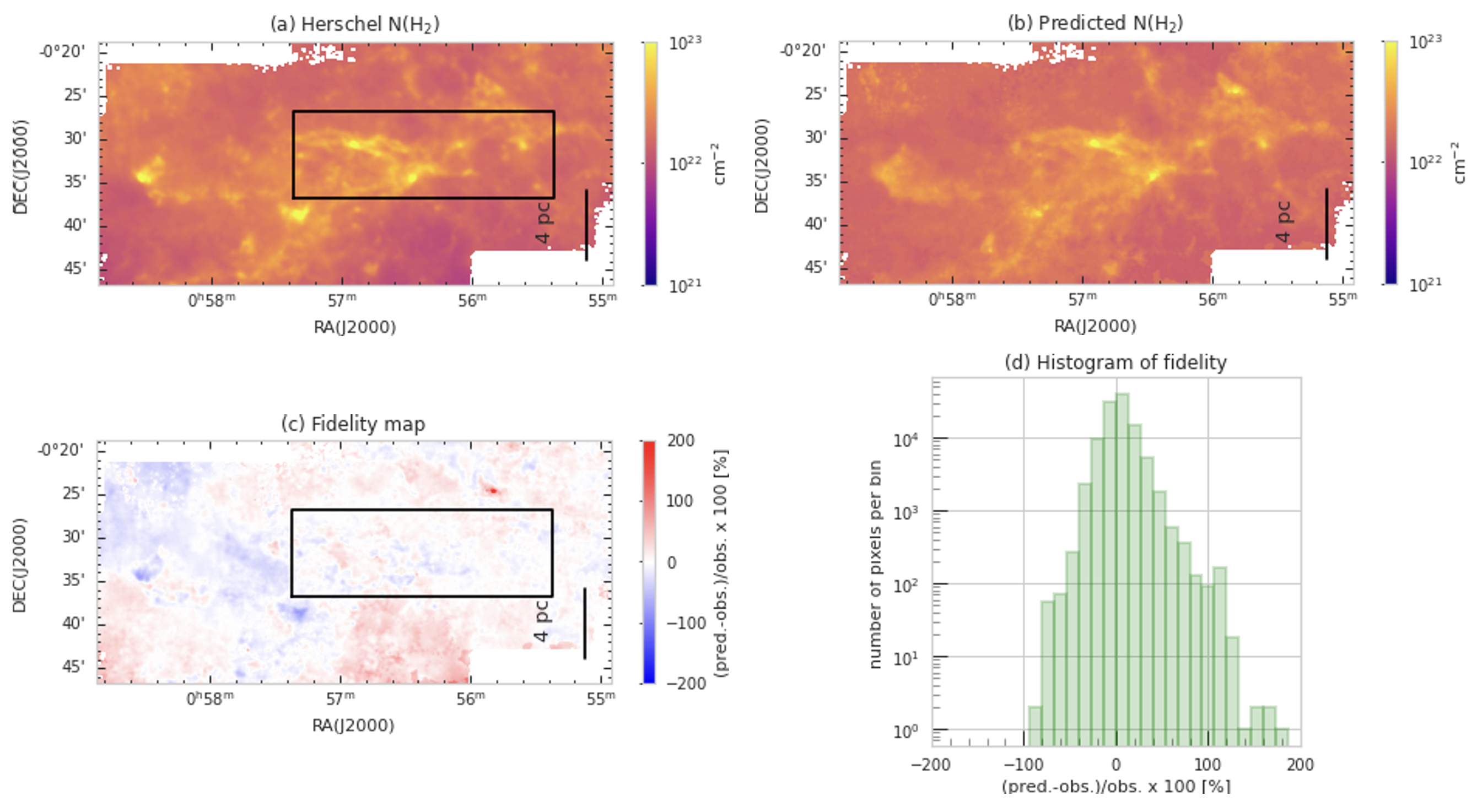}
\caption{Same as Figure \ref{fig_OMC1_prediction}, but toward M~17 and predicted by Regressor-M17.
}
\label{fig_M17_prediction}
\end{figure*}

Figure \ref{fig_OMC123_prediction} compares the predicted and observed H$_2$ column density maps of Orion~A. For the training, the data around an area encompassing OMC-1/2/3 regions is used. 
Figure \ref{fig_ngc2024_prediction} compares the predicted and observed H$_2$ column density maps of Orion~B/NGC~2024. The column density is predicted by Regressor-NGC202.  
Figure \ref{fig_aquila_prediction} compares the predicted and observed H$_2$ column density maps of Aquila. The column density is predicted by Regressor-Aquila.  
Figure \ref{fig_M17_prediction} compares the predicted and observed H$_2$ column density maps of M~17. The column density is predicted by Regressor-M17.  

\bsp	
\label{lastpage}
\end{document}